\documentclass[reprint,amsmath,amssymb,superscriptaddress,aps,pra]{revtex4-2}

\usepackage{graphicx}
\usepackage{dcolumn}
\usepackage{bm}
\usepackage{physics}
\usepackage{minitoc}

\usepackage{tocloft}

\usepackage[usenames, dvipsnames]{color}
\usepackage{enumerate}  
\usepackage{float}
\usepackage{xargs}  
\usepackage{setspace}
\usepackage[export]{adjustbox}
\usepackage{fontawesome}
\usepackage{xcolor, soul}
\sethlcolor{yellow}

\usepackage{upgreek}

\usepackage[utf8]{inputenc}

\date{\today}

\begin{document}
\doparttoc 
\faketableofcontents 

\part{} 

\title{The effect of measurement backaction on quantum clock precision studied with a superconducting circuit}

\author{Xin He}
\affiliation{ARC Centre of Excellence for Engineered Quantum Systems, School of Mathematics and Physics, The University of Queensland, St Lucia, Australia.}

\author{Prasanna Pakkiam}
\affiliation{ARC Centre of Excellence for Engineered Quantum Systems, School of Mathematics and Physics, The University of Queensland, St Lucia, Australia.}

\author{Adil A. Gangat}
\affiliation{ARC Centre of Excellence for Engineered Quantum Systems, School of Mathematics and Physics, The University of Queensland, St Lucia, Australia.}
\affiliation{Physics \& Informatics Laboratories, NTT Research, Inc., Sunnyvale, California 94085, USA}
\affiliation{Division of Chemistry and Chemical Engineering, California Institute of Technology, Pasadena, California 91125, USA}

\author{Michael J. Kewming}
\affiliation{School of Physics, Trinity College Dublin, College Green, Dublin 2, Ireland}

\author{Gerard J. Milburn}
\affiliation{ARC Centre of Excellence for Engineered Quantum Systems, School of Mathematics and Physics, The University of Queensland, St Lucia, Australia.}

\author{Arkady Fedorov}
\email[To whom correspondence should be addressed: ]{a.fedorov@uq.edu.au}
\affiliation{ARC Centre of Excellence for Engineered Quantum Systems, School of Mathematics and Physics, The University of Queensland, St Lucia, Australia.}

\begin{abstract}
We theoretically and experimentally study the precision of a quantum clock near zero temperature, explicitly accounting for the effect of continuous measurement. The clock is created by a superconducting transmon qubit dispersively coupled to an open coplanar resonator. The cavity and qubit are driven by coherent fields, and the cavity output is monitored with a quantum-noise-limited amplifier. 
When the continuous measurement is weak, it induces persistent coherent oscillations (with fluctuating periods) in the conditional moments of the qubit's energy probability distribution, which are manifest in the output of the resonator. 
On the other hand, strong continuous measurement leads to an incoherent cycle of quantum jumps. 
We theoretically find an equality for the precision of the clock in each regime. Independently from the equalities, we derive a kinetic uncertainty relation for the precision, and find that both equalities satisfy this uncertainty relation.
Finally, we experimentally verify that our quantum clock obeys the kinetic uncertainty relation for the precision, thus making an explicit link between the (kinetic) thermodynamic behavior of the clock and its precision, and achieving an experimental test of a kinetic uncertainty relation in the quantum domain.

\end{abstract}

\maketitle

\section{\label{intro}Introduction}

Quantum clocks are of importance in both technological applications and fundamental studies.  Small-scale quantum clocks will be required for deep integration into future quantum technologies such as gate-based quantum computers, where accurate clocks are required for turning control pulses on and off \cite{Ball,PhysRevApplied.12.044054}.  Atomic clocks, as a type of quantum clock, are used in searching for variations of the fundamental constants \cite{safronova2019search}.  Furthermore, the interplay between quantum clocks and gravitational redshifts may provide insights into quantum gravity \cite{JunYe}.  Creating and improving these types of periodic quantum clocks requires an understanding of their fundamental limits.

In order for a dynamical variable of a quantum clock to produce a \textit{classical} timing signal, it must undergo measurement. Because of the collapse postulate of quantum mechanics, this measurement is a fundamental source of noise in the timing signal. The objective of the present work is to shed light on how the measurement process fundamentally affects the precision of the classical timing signal produced by quantum clocks. This requires an experimental setup that allows us to study such measurement backaction \textit {in isolation} from other possible contributions to the noise in the timing signal.  The particular superconducting circuit clock that we employ (described below) meets this requirement.  While other quantum clocks, such as atomic clocks, are far more accurate than the clock that we employ, they do not meet this requirement because their complexity introduces many sources of noise into the timing signal other than the noise from measurement backaction.

To study the fundamental limits of periodic clocks, it is useful to understand them as the union of two subsystems: a driven nonlinear mechanism (system) and a counting register (measurement) that is coupled to the nonlinear mechanism \cite{Erker}.  
When the nonlinear mechanism is a quantum system, the clock is considered a quantum clock; otherwise, it is a classical clock.  
The counting register corresponds to the classical measurement outcomes of the nonlinear system.  
For the nonlinear mechanism to be measured continuously, it is necessarily dissipative.
Furthermore, it evolves cyclically in time; each cycle corresponds to a single tick of the clock. 
The degree of regularity of these ticks in the absence of coupling to the counter is referred to as the \textit{accuracy} of the clock.
The counter of the clock registers the ticks of the clock to produce a timing signal.  
The degree of regularity of the timing signal is referred to as the \textit{precision} of the clock.  
Since coupling the nonlinear mechanism to the counter necessarily affects the dynamics of the former, the accuracy and precision of a clock are not the same.

In considering the fundamental limits to clock performance, it has been recognized that all clocks are constrained by the laws of thermodynamics that limit their accuracy and precision~\cite{Milburn-CP}.  However, for thermodynamic systems, including clocks, that operate far from equilibrium, \textit{kinetic} considerations also play a fundamental role \cite{Maes_frenetic_2017, Di_Terlizzi_2018}.
Somewhat surprisingly, both thermodynamic and kinetic considerations of the fundamental limits to classical and quantum clock accuracy and precision have been neglected until only recently \cite{Erker,Woods_2022,Mitchison,PhysRevE.105.055001, Pearson, Woods,Schwarzhans_Autonomous_2021}. Arguably the most important development along these lines thus far is that a thermodynamic bound has been derived on the accuracy of generating clock ticks \cite{Meier_2023}. Still missing in this nascent program of research is a consideration of clock \textit{precision} from a kinetic point of view.  In this work, we address this gap by theoretically and experimentally studying the fundamental kinetic constraints on the precision of a particular quantum clock.  As mentioned, this necessarily includes a consideration of contributions from measurement backaction.

For a quantum clock to generate a (classical) timing signal, it must contain both a quantum system (i.e., the nonlinear mechanism) and a (classical) measurement apparatus (i.e., the counter) that monitors the quantum system.  
If the conditional dynamics (i.e. the dynamics that includes the effect of monitoring) of the quantum system is periodic, the manifestation of this periodicity in the classical measurement record serves as the timing signal.  
Because of the continuous measurement, the period ($T$) of the timing signal will be subject to fluctuations induced by quantum noise and measurement backaction \cite{hatridge2013quantum}, even at zero temperature.
Therefore, we need to consider the average period (${\rm E}[T]$) and the variance in the period (${\rm Var} [T]$). 
We define the precision \footnote{In Ref. \cite{Erker}, $N$ is referred to as accuracy because $T$ in their study is the period of the ticks, not the period of the timing signal.} of the clock as 
\begin{equation}
\label{eq:precision}
    N=\frac{{\rm E} [T]^2}{{\rm Var} [T]}.
\end{equation}

Studies of particular theoretical models or experimental systems can lead to fundamental insights that apply beyond the particular model or system.  For gaining fundamental insights into the effect of continuous measurement on quantum clock precision, in this work we specifically consider a coherently driven qubit subject to continuous measurement (i.e. monitoring) at very low temperatures \footnote{We note that any clock, with a suitable redefinition, can be considered \textit{autonomous}, including the one in the present study. But, in contrast to the case in Ref. \onlinecite{Erker}, this conceptual framework is not useful for our purposes here, so we do not employ it.}.  The platform that we use (described below) has the advantage (compared to, for example, atomic clocks) of enabling a measurement of the quantum system with very low noise such that we are able to study in isolation the effect of quantum measurement backaction on clock precision.
While the driven and unmonitored qubit necessarily relaxes into a stationary state due to decoherence, the driven and weakly monitored qubit acquires a \textit{persistent} dynamical state:
weak continuous measurement continually restores quantum coherence and thereby enables the coherent drive to create persistent, but noisy, Rabi oscillations~\cite{Wiseman_stochastic_1993, goan2001dynamics,korotkov2001continuous,palacios2010experimental}.
Under strong monitoring, the coherently driven qubit operates as an incoherent quantum jump clock: the Zeno effect pins the qubit to an eigenstate while quantum and thermal noise occasionally lead to a switch of that eigenstate~\cite{vijay2011observation}.

To understand how quantum measurement factors into the (kinetic) thermodynamics of precision of a quantum clock, one must consider the recent developments in thermodynamic uncertainty relations (TURs).
These relations were initially developed to describe the trade-off between precision and dissipation in time-integrated currents for biomolecular processes, and classical Markov processes~\cite{Barato_2015, Gingrich_2016, Horowitz_2017, Dechant_2018, Liu_thermodynamic_2020}, but were also used to describe the behavior of thermodynamic heat engines \cite{Shiraishi_universal_2016, Pietzonka_universal_2018}, and inference of thermodynamic quantities \cite{Junang_quantifying_2019, vu_entropy_2020, Sreekanth_2020, otsubo_estimating_2020}.
TURs place a thermodynamic bound on the precision by the entropy production and have an incredibly wide range of applications.
This has led to, in recent years, a rapidly growing body of research into the thermodynamics of precision and extending TURs to many classical and quantum examples \cite{Brander_thermodynamic_2018, Agarwalla_assessing_2018, Ptaszy_coherence_2018, Saryal_thermodynamic_2019, Vu_uncertainty_2019, Liu_thermodynamic_2019, Hasegawa_uncertainty_2019, Timpanaro_thermodynamic_2019, Guarnieri_thermodynamics_2019, horowitz_2020_thermodynamic, Vu_thermodynamic_2020, Koyuk_thermodynamic_2020, Potts_thermodynamic_2019, Vo_unified_2020, falasco2020unifying, Friedman_thermodynamic_2020, Cangemi_violation_2020,  Falasco_dissipation_2020, Kalaee_violation_2021, Sacchi_thermodynamic_2021, Park_thermodynamic_2021, Lee_quantumness_2021, Dechant_continuous_2021, Miller_thermodynamic_2021, Rignon-Bret_2021, Timpanaro_2021_1, menczel2021thermodynamic,kewming2022entropy, Ray_2022, Prech_2022}.

Little is known about the application of uncertainty relations to the case of \textit{clock} precision. One quantity of interest for such a relation is the first passage time (FPT). The FPT is the average time required for any cyclic system to undergo a single cycle. This quantity is, therefore, of very broad relevance to open quantum systems.  In the context of clock precision, the FPT corresponds to the average period of the timing signal; the variable $T$ in Eq.(\ref{eq:precision}) is the FPT for this study.
The FPT has been theoretically studied in the context of TURs \cite{Gingrich_fundamental_2017} and, more generally, in terms of kinetic uncertainty relations (KURs) \cite{Vu_thermodynamics_2022}. 
KURs, like TURs, put a bound on the precision, but in terms of the dynamical activity $\mathcal{N}$---which is a measure of the total dynamic transitions or jumps that occur in a process---instead of the entropy production. 
KURs are arguably more general than TURs, as they apply to systems far from equilibrium and to systems that do not satisfy detailed balance \cite{Maes_frenetic_2017, Di_Terlizzi_2018, Garrahan_simple_2017, di2018kinetic, Prech_2022}.

Experimental tests of KURs for the FPT (or any other quantity) remain absent in the quantum domain. Here we address this gap by using a superconducting circuit realization of a quantum clock.
We show both theoretically and experimentally how a KUR for the FPT ($T$) of our clock's timing signal applies to our clock's precision (Eq. (\ref{eq:precision})). Independently from the KUR, we derive equalities for the precision ($N$) of the clock in both the quantum coherent and incoherent regimes.  We find that these equalities satisfy the KUR; the equality for $N$ in the quantum coherent regime achieves the unbounded increase in $N$ permitted by the KUR, while the equality in the incoherent regime reveals a saturation of $N$.  Our experimental results strongly confirm the equality for $N$ in the incoherent regime, but do not match well with the equality for $N$ in the quantum coherent regime.  The discrepancy in the quantum coherent regime invites further investigation, but overall these results clearly show that the measurement process and its thermodynamic-kinetic contribution must be accounted for when studying the fundamental limits of a quantum clock.

This article is organized as follows. In Sec.~\ref{SC-clock}, we detail the theoretical modeling and experimental realization of the quantum clock comprising a transmon qubit dispersively coupled to a superconducting coplanar waveguide. We discuss in detail how one can operate this clock in both the weak measurement ``oscillatory'' regime and the strong measurement ``jump'' regime. In Secs. \ref{sec:coherent_equality} and \ref{sec:jump_equality}, we discuss how one can obtain the theoretical equalities for the statistics of the clock (and hence its precision) for both regimes by using the FPT picture, and we present experimental validation of this modeling. In Sec.~\ref{sec:thermodynamic_clocks}, we turn to deriving the KUR for the FPT from the Cramer-Rao bound, which leads to an inequality for the clock precision, and we show experimental confirmation of this inequality in both clock regimes. We end by summarizing our results and indicating future directions in Sec. \ref{sec:conclusion}.

\section{\label{SC-clock}A superconducting quantum clock}

\begin{figure}
    \centering
    \hspace*{-0.02in}
     \includegraphics[width=1\columnwidth]{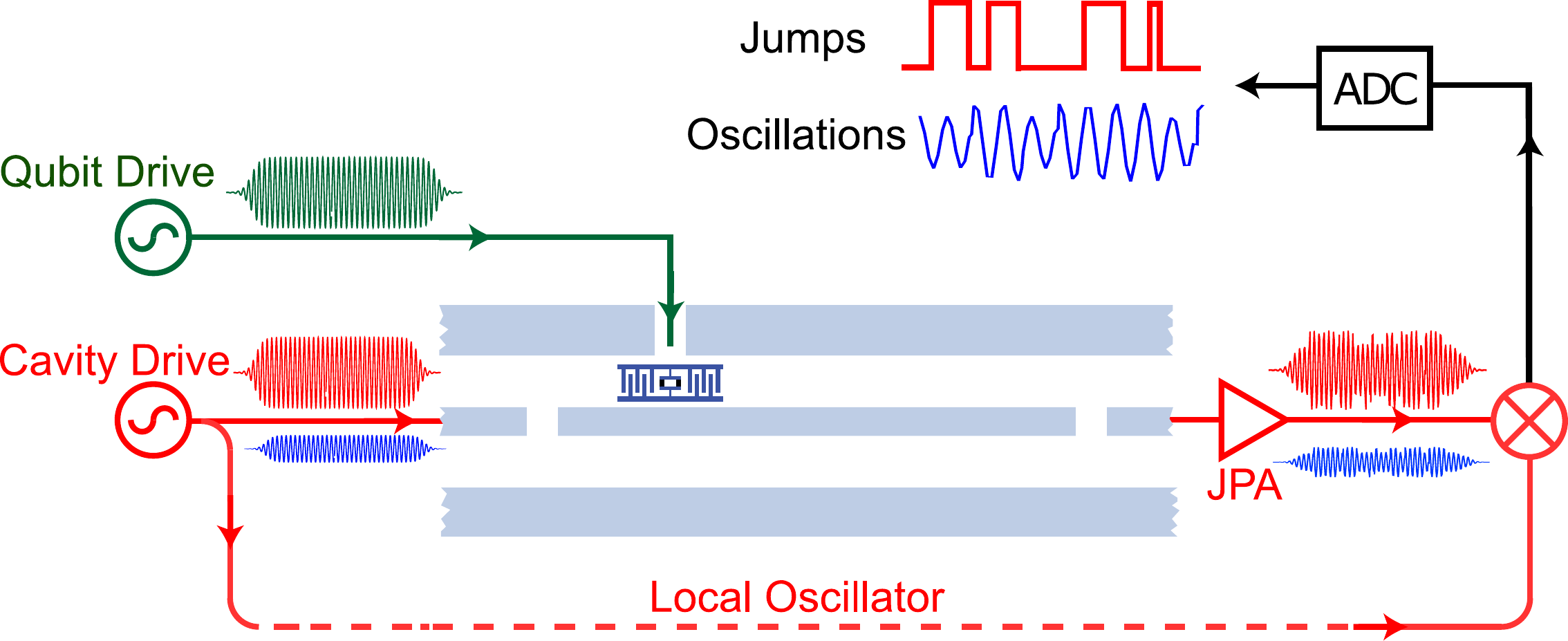}
    \caption{A model of the homodyne measurement scheme used to extract the clock signal from the source cavity (see Eq.~(\ref{eqn:EOM_homodyne_main})). A transmon qubit is dispersively coupled to a superconducting microwave cavity. Both the cavity and qubit are driven coherently. The output field is subject to a homodyne measurement, resulting in conditional dynamics of the cavity field that, with signal acquisition and processing using an analog-to-digital converter (ADC) and filtering, produces the clock signal for counting. The actual experiment employs a quantum-noise-limited Josephson parametric amplifier (JPA), but the resulting conditional dynamics of the qubit itself is the same as under the conventional homodyne scheme represented here. As explained in the text, a smaller amplitude cavity drive (dark blue) results in sustained oscillations at the ADC output, while a larger cavity drive amplitude (red) results in jumps. These cycles of oscillations or jumps reflect the dynamics of the qubit, and thereby serve as ticks of a quantum clock; the precision of these quantum clock ticks is what we investigate.}
    \label{main:homodyne-scheme}
\end{figure}

We theoretically model and experimentally investigate a quantum clock comprising a superconducting circuit and measurement circuitry.  As illustrated in Fig.~\ref{main:homodyne-scheme}, the superconducting circuit consists of a superconducting coplanar waveguide resonator (``cavity") dispersively coupled to a transmon qubit. The cavity and the qubit are coherently driven on resonance. The coupling of the driven cavity to the qubit enables a measurement channel via the phase of the emitted cavity field, to monitor the evolution of the qubit's population in the energy eigenbasis.  The population evolution is either oscillatory or jumplike, depending on the strength of the continuous measurement.  
 
 This is not a good quantum clock for technological applications, as its parts are not amenable to microscale integration. However, it is a good quantum clock for the purpose here of the fundamental study of quantum clock precision because it provides the requisite amount of control and measurement fidelity (as demonstrated with the experimental data below).

 In the experiment, the transmon qubit is tuned to have a frequency of $4.75\,\mathrm{GHz}$ (see Appendix \ref{sec:qubit_tuning} for details), and the longitudinal ($T_1$) and transverse ($T_2$) relaxation times of the qubit are respectively $T_1=25\,\upmu\mathrm{s}$ and $T_2=3\,\upmu\mathrm{s}$. The dispersive shift ($\chi/2\pi$) is $340\,\mathrm{kHz}$. The microwave resonator frequency $(\omega_{c}/2\pi)$ is $ 7.02\,\mathrm{GHz}$, and its loaded quality factor is 9623, corresponding to a total cavity linewidth ($\kappa/2\pi$) of $0.73\,\mathrm{MHz}$.

The circuit chip is cooled by a dilution refrigerator down to $22\,\mathrm{mK}$. 
We drive and measure the qubit via microwave signals according to details in Appendix~\ref{appen:expsetup}. Our scheme gives direct access to the conditional cavity field quadrature $\langle \hat{x}(t)\rangle_c$ (where $\hat{x}=a+a^\dagger$ with $a$ the cavity field annihilation operator and the subscript denoting a quantity that is conditional on the measurement outcomes) with minimal added noise because of the use of a Josephson parametric amplifier (JPA).

We model the transmon qubit as a two-level system subject to very weak spontaneous emission, and the cavity as a simple harmonic oscillator with amplitude damping, the dominant dissipative mechanism in the system. The conditional state of the total field-qubit system, in a frame rotating at the frequencies of the cavity drive and qubit drive, can be described by the stochastic master equation for homodyne detection ($\hbar=1$):
\begin{eqnarray}
\label{eqn:EOM_homodyne_main}
   d\rho_c &=&  -iE[a+a^{\dagger},\rho_c]dt -i\Omega[\sigma_x,\rho_c]dt \\\nonumber &&-i\chi[a^{\dagger}a\sigma_z,\rho_c]dt+\gamma{\cal D}[\sigma_-]\rho_c+ \kappa{\cal D}[a]\rho_c dt\\ \nonumber
    && +\sqrt{\eta\kappa}{\cal H}[a]\rho_c dW(t).
\end{eqnarray}
Here $dW(t)$ is a Wiener processes, $E$ is the cavity drive amplitude, $\Omega$ is the qubit drive amplitude, $\gamma$ is the qubit amplitude damping rate, $\kappa$ is the cavity damping rate through the output port, $\eta$ is the efficiency of the photodetector, and $\mathcal{D}[A]$ and $\mathcal{H}[A]$ are superoperators: ${\cal D}[A]\rho  =  A\rho A^\dagger-\frac{1}{2}(A^\dagger A \rho+\rho A^\dagger A)$ and ${\cal H}[A]\rho  =  A\rho+\rho A^\dagger -{\rm tr} (A\rho +\rho A^\dagger)\rho$.  For completeness, we note that by averaging the conditional master equation over the noise, we obtain the {\em unconditional} dynamics.

In the actual experiment, we used a JPA in the nondegenerate mode and heterodyne detection scheme, which should produce similar conditional qubit dynamics but with larger noise. In the limit of an ideal quantum-limited amplifier, the noise will be twice larger~\cite{Eichler2012} and less than twice for the actual experiment.  The derivation of the stochastic master equation for the heterodyne detection will be published in future work.   
Because of the dispersive coupling between the cavity and qubit, the phase of the cavity field is dependent on the state of the qubit; continuously measuring a quadrature of the cavity output serves as a way to continuously measure the qubit observable $\sigma_z$. A stronger drive on the cavity leads to more information about the qubit leaving the cavity per unit time (i.e. a stronger continuous measurement), and a weaker cavity drive leads to a weaker continuous measurement.

For sufficiently weak continuous measurement, it is known that this can result in persistent (but noisy) Rabi oscillations of the qubit \cite{goan2001dynamics,korotkov2001continuous,palacios2010experimental}. The noise in the oscillations is fundamentally due to quantum noise in $\sigma_z$, and measurement backaction \cite{hatridge2013quantum}.
Thus, the conditional dynamics of the qubit under continuous measurement of $\sigma_z$ yields the continual ticking of the clock, though with a necessarily finite precision because of fundamental noise.

When the continuous measurement strength ($\Gamma$) becomes sufficiently \textit{large}, a measurement-induced suppression (i.e. quantum Zeno effect) occurs on the qubit oscillations. 
This would appear to disable the clock, but in fact, this gives rise to a very different kind of noisy clock: a \textit{quantum jump} clock. That is, it is known \cite{vijay2011observation} that in this regime of operation, the competition between the quantum Zeno effect and the qubit drive gives rise to an incoherent switching of the qubit state that results in a random telegraph process. 

We are primarily interested in the dynamics of the qubit and not the readout cavity.
We, therefore, adiabatically eliminate the cavity field (see Appendix \ref{sec:adiabatic_elimination}), which leads to a conditional master equation for the reduced state of the qubit ($\rho_{\sigma}$) only:
\begin{align}
\label{eq:qubit_me}
   d\rho_\sigma &= - i[H_{\sigma},\rho_\sigma]dt
   +\gamma{\cal D}[\sigma_-]\rho_\sigma dt
   \nonumber \\ 
   &\qquad+\Gamma{\cal D}[\sigma_z]\rho_\sigma dt
   -\sqrt{\Gamma}{\cal H}[\sigma_z]\rho_\sigma dW(t).
\end{align}
Here $H_\sigma = \Omega \sigma_{x} + \Delta \sigma_{z}$ is the Hamiltonian acting on the qubit, $\Delta = \chi|\alpha_0|^2$ is the effective Stark shift of the qubit due to the dispersive interaction with the cavity, $\Gamma=4\chi^{2}|\alpha_0|^2/\kappa$ is the dephasing rate of the qubit due to measurement backaction, and
$\alpha_{0} = -2iE/\kappa$ is the coherent state amplitude of the cavity.
Though $\kappa$ and $\chi$ are experimentally fixed, as described above, $\Gamma$ can be tuned by varying $E$. Thus, the single superconducting circuit that we have can be tuned between the oscillatory and jump clock modes simply by changing $E$.

Below we present experimental observations of the oscillatory and quantum jump clock signals using the aforementioned setup. The experimental results show good agreement with QuTip \cite{johansson2012qutip,johansson2013qutip} simulations (shown in Appendix \ref{sec:theory}) of the model represented by Eq.~(\ref{eqn:EOM_homodyne_main}).

\subsection{\label{sec:coherent_equality}Rabi oscillation regime: experimental oscillatory clock}

As mentioned, we enter the oscillatory clock regime by tuning the cavity drive amplitude ($E$) to be sufficiently weak.  However, a weak cavity drive leads to a small (average) intracavity photon number, which is generally in conflict with high signal-to-noise ratio (SNR) measurements. This is compensated for in the experiment by carefully tuning the JPA frequency and gain to have the maximal SNR at each given Rabi oscillation frequency.

\begin{figure}
\centering
\hspace*{-0.3in}
\includegraphics[width=1.15\columnwidth]{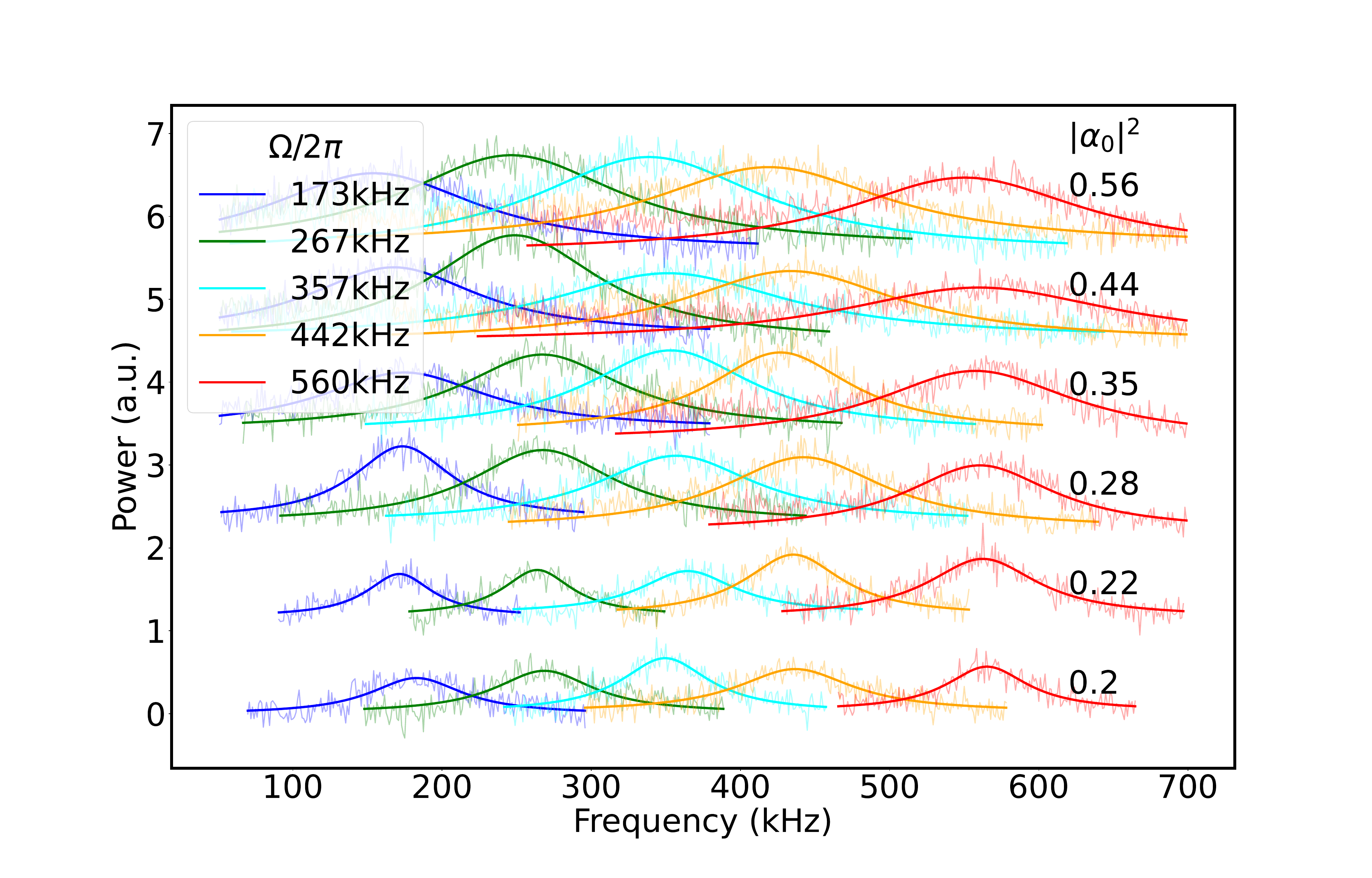}
\caption{Oscillatory clock noise-floor-subtracted power spectral density (PSD) traces of the conditional cavity quadrature dynamics ($\langle a + a^{\dagger}\rangle_c(t)$). The PSD is shown as a function of qubit drive amplitude ($\Omega$) and readout resonator average photon number ($\vert\alpha_0\vert^2$, labeled right). The faint signals are the PSD data, while the solid lines indicate fitted Lorentzians.  Larger $\vert\alpha_0\vert^2$ leads to stronger measurement of the qubit (i.e., larger $\Gamma$) and, therefore, more phase noise in its Rabi oscillations, and hence broader PSD traces.
}
\label{fig:PSDs}
\end{figure}

We first test our oscillatory clock by investigating six different drives $E$ at five different values of $\Omega$.  The Rabi oscillations of the qubit modulate the readout resonator frequency, leading to oscillations in $\langle a+a^{\dagger}\rangle_c(t)$ and a sideband of the readout resonator output signal. By mixing down the signal with the readout resonator drive tone and taking the power spectral density (PSD), the Rabi oscillation signals are observed in the spectra as shown in Fig.~\ref{fig:PSDs}. 

Each PSD trace in Fig.~\ref{fig:PSDs} is obtained from the fast Fourier transform of a 400-ms-long time trace of $\langle a+a^{\dagger}\rangle_c(t)$, and then processed by removing the JPA-dominated noise floor. In this figure, $\Omega$ is changed within each row, and the Rabi oscillation frequency changes correspondingly.
The resonator intracavity average photon number ($\vert \alpha_0 \vert^2$) varies by row, and is labeled at the right. Decreasing the resonator photon number (from top to bottom rows in Fig. \ref{fig:PSDs}) causes a narrowing of the Rabi peaks, which is attributed to the decrease of the measurement-induced dephasing \cite{Szombati2019},  
whereas the Stark shift is negligible, as can be seen in Fig.~\ref{fig:PSD_freqs} in Appendix~\ref{section_experimental_details}.
However, the amplitude of the Rabi peaks also decreases with decreasing resonator photon number, which is due to the degradation of SNR. 
Given that we want a maximum SNR, we find the optimal readout resonator average occupation at $\vert \alpha_{0}\vert^{2} =0.28$ photons, and we use the corresponding value of $E$ for the rest of the experimental investigation of the oscillatory clock.

By fitting the Lorentzian function to the PSD peaks that are obtained with the optimal value of $E$, the Rabi oscillation frequencies are extracted and plotted as a function of qubit drive amplitude in Fig.~\ref{fig:PSD_freqs} in Appendix B, which confirms that the oscillation frequency is linearly dependent on the qubit drive amplitude ($\Omega$) over the range in Fig.~\ref{fig:PSDs}; this linear dependency agrees with that which is independently obtained with the qubit drive calibration that is done in the time domain with a different set of $\Omega$ (see Fig. \ref{fig:2D_rabi} in Appendix B).

\begin{figure}
\centering
\hspace*{-0.3in}
\includegraphics[width=1.15\columnwidth]{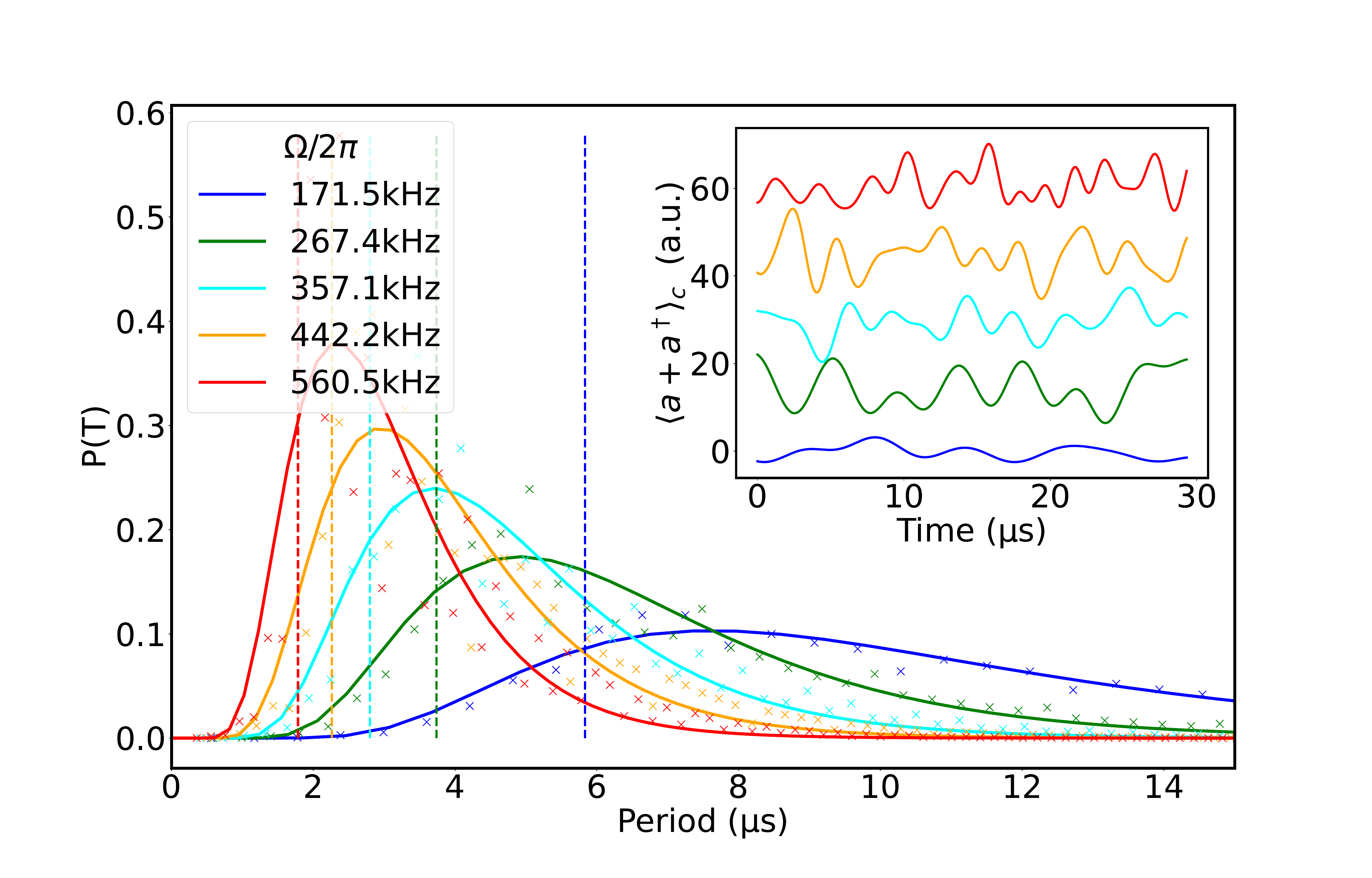}
\caption{Oscillatory clock period distributions (main figure) and time-domain signal (inset). The main figure has normalised histograms for the experimental data (points) and theoretical fits (solid lines) of inverse Gaussians [Eq. (\ref{eq:Wald})] to the data at five different qubit drive amplitudes ($\Omega$). The experimental data are from the time-domain analysis described in the main text. The colour scheme denoting the different $\Omega$ is the same as in Fig. \ref{fig:PSDs}.  As seen by the data and the fits, increasing $\Omega$ causes the measured signal to oscillate faster with a tighter spread in its periods.
}
\label{fig:wald_fits}
\end{figure}

\begin{figure}
    \centering
    \includegraphics[width=\columnwidth]{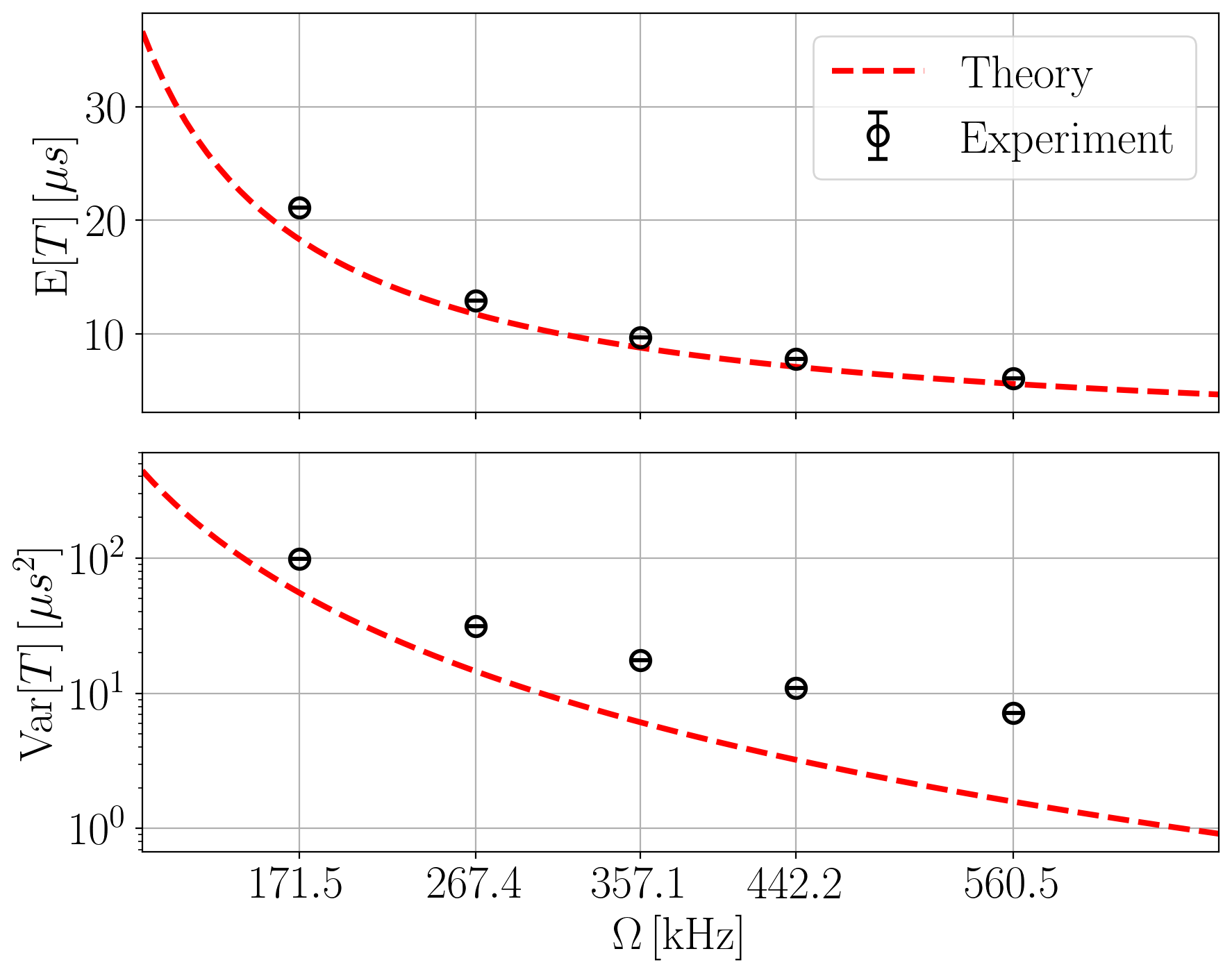}
    \caption{Oscillatory clock period ($T$) statistics as a function of qubit drive amplitude ($\Omega$) From top to bottom: mean (${\rm E} [T]$), variance (${\rm Var}[T]$,), and precision ($N$). The theoretical curves are from Eq. (\ref{eq:N_osc}), corresponding to an inverse Gaussian distribution. The experimental estimates of ${\rm E} [T]$ and ${\rm Var}[T]$ are computed directly from the data for each $\Omega$. The larger experimental variance than predicted by the theory leads to a significant decrease in the experimental $N$ from the theoretical prediction.}
    \label{fig:wald_data}
\end{figure}

After confirming the Rabi oscillation signals in the frequency domain, we use the time traces (one for each value of $\Omega$) for $\langle a+a^{\dagger}\rangle_c(t)$ that were obtained with the optimal value of $E$ to evaluate the clock statistics in the time domain. For each value of $\Omega$, after some preprocessing of the time-trace data (see Appendix \ref{sec:pre-processing}), the clock tick signal is generated from applying the sign() function to the trace after passing it through a low-pass filter with the cutoff frequency equal to the Rabi frequency. In the resulting telegraph signal, the length between two \textit{rising} zero crossings is the clock period (i.e. FPT).

On the theoretical side, we show in Appendix \ref{sec:Oscillatory Dynamics} that the approximate equations of motion in this regime yield the following inverse Gaussian probability density \cite{Aminzare} for the FPT distribution:
\begin{equation}
\label{eq:Wald}
    P(T)= \sqrt{\frac{\pi}{\Gamma T^3}}\exp\left [-\frac{(\pi-\Omega T)^2}{\Gamma T}\right ]\,.
\end{equation}
This distribution gives values of E[$T$], Var[$T$], and the precision as
\begin{equation}
\label{eq:N_osc}
   {\rm E}[T] = \frac{\pi}{\Omega} \,,\quad {\rm Var}[T] = \frac{\pi\Gamma}{2\Omega^3} \,, \quad  N_{\rm osc} = 2\pi \left(\frac{\Omega}{\Gamma}\right)\,.
\end{equation}
Naturally, as the Rabi frequency increases, this coincides with a shorter expected tick time corresponding to a clock with higher resolution. Clearly, however, the precision of the clock ($N_{\rm osc}$) decreases for an increasing $\Gamma$, indicating the deleterious effect of the measurement backaction.

We first verify that the experimentally obtained clock period from the time-domain data does obey an inverse Gaussian distribution.  This is shown in Fig.~(\ref{fig:wald_fits}) by the qualitatively good fits of Eq. (\ref{eq:Wald}) to the time-domain data at five different values of $\Omega$. However, the experimental time-domain signals provide us with a direct measurement of the mean (E[$T$]) and variance (Var[$T$]) of the period (i.e. FPT) for this clock, which we use instead of extracting E[$T$] and Var[$T$] from the fits.

At the five different values of $\Omega$, we compare these experimental and theoretical results for E$[T]$ and Var$[T]$ in Fig. \ref{fig:wald_data} using the calibrated parameters in the theoretical models.  We find agreement in the monotonic behavior and the order of magnitude.  Thus we conclude that in the oscillatory regime our model of such a clock is experimentally validated. However, as seen in the middle panel of Fig. \ref{fig:wald_data}, there are clearly systematic errors that are unaccounted for in the variance.  This leads to a significantly lower experimental oscillatory clock precision ($N$) than the theoretical prediction ($N_{osc}$) in Eq. (\ref{eq:N_osc}), as seen in the bottom panel of Fig. \ref{fig:wald_data}. 

\begin{figure}
\centering
\hspace*{-0.15in}
\includegraphics[width=1.1\columnwidth]{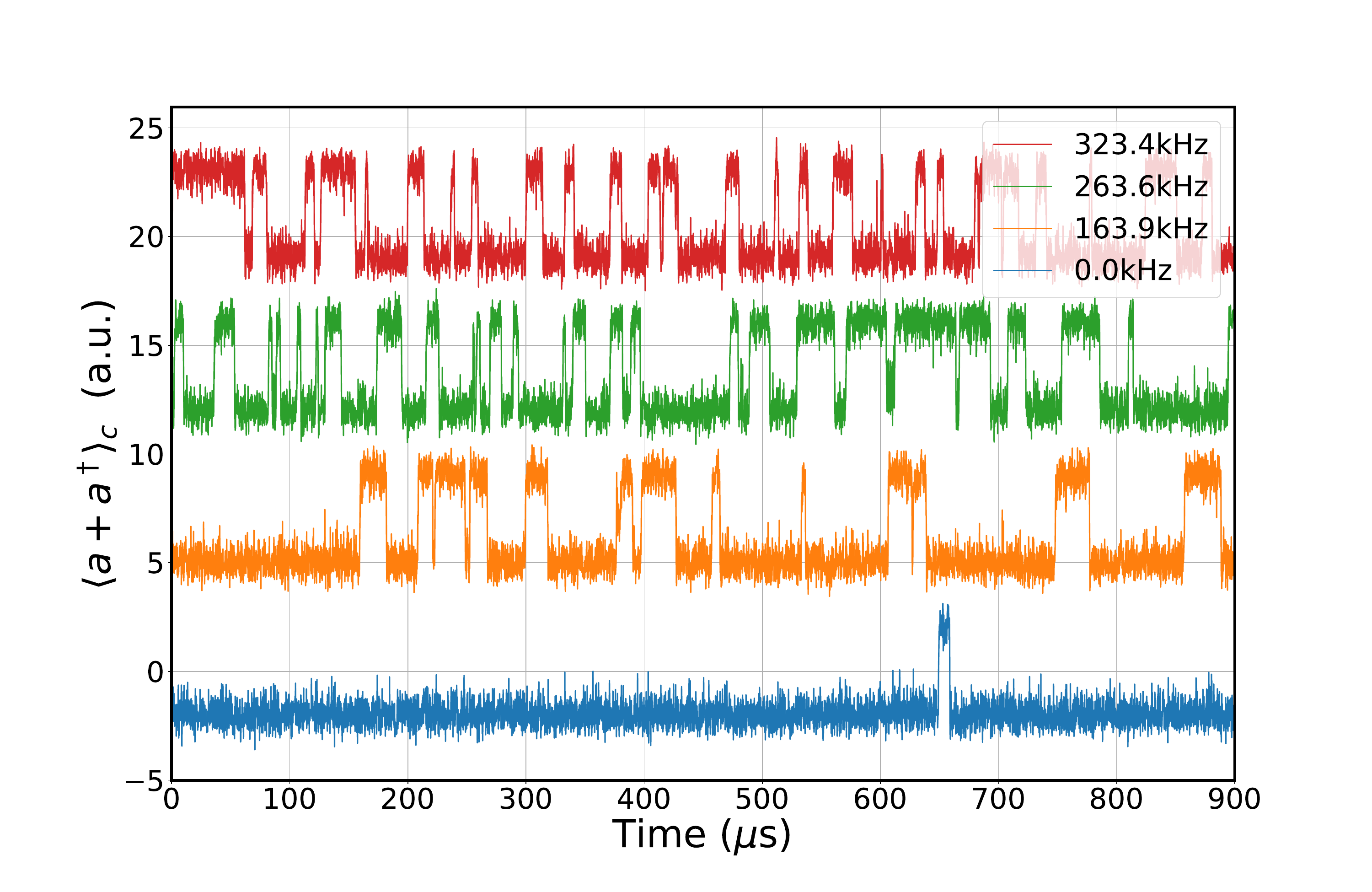}
\caption{Observed quantum jumps in the strong measurement regime. Sample of the time traces indicating quantum jumps (between ground and excited states) with different qubit drive amplitudes ($\Omega$, shown as different colours with different vertical offsets). In each curve, a low signal indicates the ground state, and a high signal indicates the excited state. Thermal noise yields a small excited state population when using zero qubit drive. As the qubit drive amplitude increases, the qubit approaches an equal amount of time spent in the ground and excited states. }
\label{fig:jumps}
\end{figure}

\begin{figure}
\centering
\hspace*{-0.3in}
\includegraphics[width=1\columnwidth]{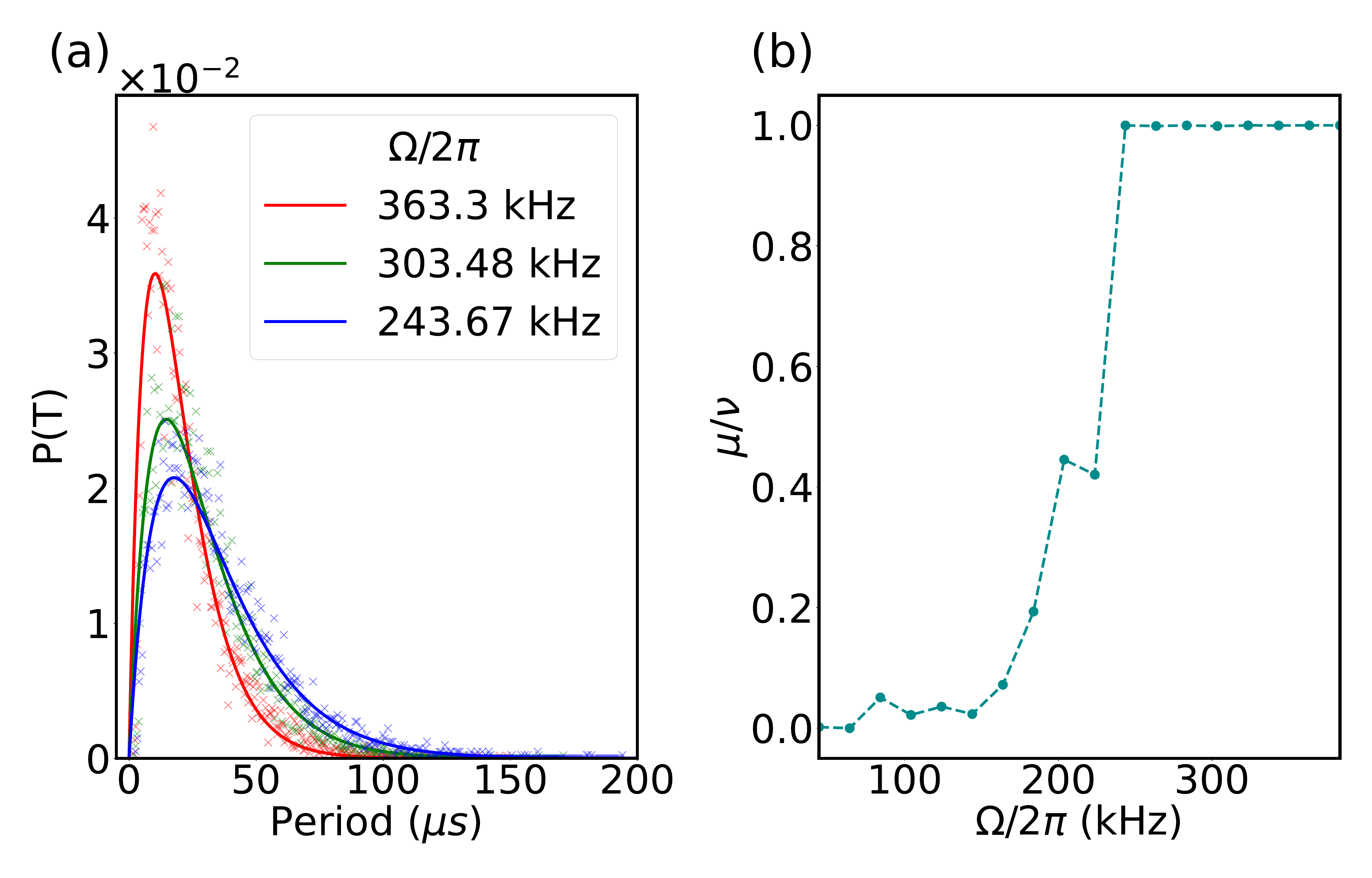}
\caption{Jump clock period ($T$) distributions and transition rate ratios ($\mu/\nu$). (a) Histograms for the experimental data (points) and theoretical fits (solid lines) of Poisson distributions [Eq. (\ref{eq:Poisson})] to the data at three different qubit drive amplitudes ($\Omega$). As in the weak measurement case (i.e oscillatory clock), increasing $\Omega$ shortens the average clock period and reduces the period variance. (b) Ratio between the upward ($\mu$) and downward ($\nu$) transition rates as a function of qubit drive amplitude. Note that the ratio becomes approximately unity at the higher qubit drive amplitudes. As the qubit spends less time in the ground state at higher qubit drive amplitudes, the upward transition rate conversely increases. The transition rates are obtained by fitting general Poisson distributions (i.e. where $\mu\neq\nu$) to the clock period histograms.
}
\label{fig:Poisson}
\end{figure}

\begin{figure}
    \centering
    \includegraphics[width=\columnwidth]{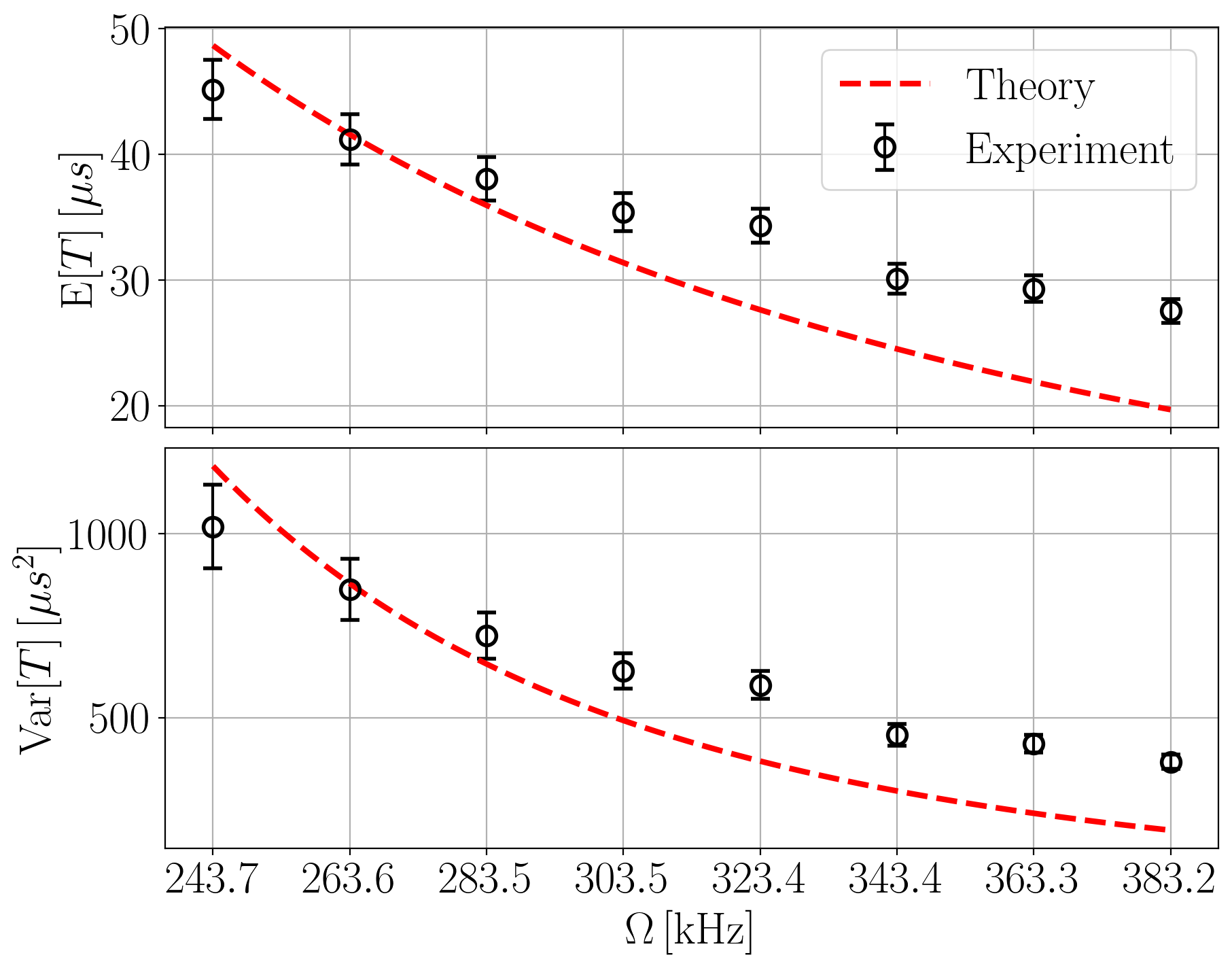}
    \caption{Jump clock period ($T$) statistics as a function of qubit drive amplitude ($\Omega$). From top to bottom: mean (${\rm E} [T]$), variance (${\rm Var}[T]$,), and precision ($N$). The experimental data at each Rabi frequency are obtained without fitting by the method described in the text. The theoretical curves are given by Eq. (\ref{eq:jump_stat1}) using the experimentally measured rates $\mu$. }
    \label{fig:jump_data}
\end{figure}

\subsection{\label{sec:jump_equality}Quantum jump regime: experimental jump clock}

As mentioned, we can enter the jump clock regime of the superconducting circuit simply by increasing the cavity drive amplitude ($E$) in the experiment. A high-gain JPA can provide enough fidelity to enable single-shot measurements \cite{Mallet2009}.  
Thus, by using a high-gain JPA for the continuous measurement, we obtain high-contrast quantum jump signals, and sections of the signals of different qubit drive amplitudes are plotted in Fig.~\ref{fig:jumps}. 
The quantum jump signals are acquired by continuously monitoring the qubit state over 250 ms. The four traces correspond to qubit drive amplitudes ranging from 0 to $323.4\,\mathrm{kHz}$. 
In order to avoid the overlap between traces, each trace is consecutively shifted by $7$ on the y axis compared to the previous one. As shown in the figure, in the lowest trace with no coherent qubit drive (i.e. $\Omega=0$), the signal amplitude stays low most of the time, showing the qubit is mostly in its ground state with one jump due to thermal noise. As $\Omega$ increases, the high-amplitude signal occurs more and more often, and the measurement signal shows very high-contrast quantum jumps. By treating each upward jump as a clock tick, the qubit in this strong measurement regime acts as a nonoscillatory clock.

We now want to theoretically estimate the statistics of the FPT (i.e., the period ($T$)) for the jump clock, which in this limit, can be approximated as a classical Markov stochastic process.  We provide details in Appendix \ref{sec:Jump Dynamics}. In this picture, the upward and downward transitions are independent, and the process is, therefore, a Poisson process. The upward rate ($\mu$) and the downward rate ($\nu$) can be different. But by fitting a Poisson distribution that allows for $\mu\neq\nu$ to the clock period histograms, we find that at sufficiently large $\Omega$, the rates are approximately equal.  This is shown in Fig. \ref{fig:Poisson}b. Therefore, we take the limit $\mu\rightarrow\nu$ and obtain the probability density 
\begin{equation}
\label{eq:Poisson}
    P(T) = \mu^{2} T e^{-\mu T}\,,
\end{equation}
where $\mu=\Gamma \Omega^{2}/(\Gamma^{2} + \Delta^{2})$ is the transition rate.  Figure \ref{fig:Poisson}a shows that Eq. (\ref{eq:Poisson}) fits the jump clock period distributions well at sufficiently large values of $\Omega$.
This distribution respectively has a mean, variance, and precision given by
\begin{equation}
\label{eq:jump_stat1}
   {\rm E}[T] = \frac{2}{\mu}, \quad {\rm Var}[T]=\frac{2}{\mu^{2}}, \quad N_{\rm jump}= 2\,.
\end{equation}
Thus, our jump clock in the regime of sufficiently large $\Omega$ has a theoretical precision that is constant, in contradistinction to the case of the oscillatory clock (Eq. (\ref{eq:N_osc})).
Experimentally, we fix the cavity drive in the strong measurement regime and then only include qubit drive amplitudes $\Omega/2\pi > 243\,{\rm kHz}$, where the upward ($\mu$) and downward ($\nu$) transition rates are approximately equal (as shown in Fig.~\ref{fig:Poisson}b) so that the statistics of the jump clock are well approximated by Eq.~(\ref{eq:jump_stat1}). 
By using the measured values of $\mu$ and the given values of $\Omega$, we use the expression for $\mu$ given below Eq.~(\ref{eq:Poisson}) to estimate $\Gamma \approx 5.8(3) {\rm MHz}$ and  $\Delta \approx 2.3(1) {\rm MHz}$.

To proceed to experimentally verify Eq. (\ref{eq:jump_stat1}), we obtain experimental values for E$[T]$ and Var[$T$] from the experimental data for $\langle a + a^{\dagger}\rangle_c(t)$ in the time domain, using a single trace of length 250 ms at each of eight values of $\Omega$. As in the oscillatory clock regime, we apply low-pass filtering, but this time with a cutoff frequency of $2.5\,\Omega$, and then perform the $\textrm{sign()}$ operation.  This yields a binary clock signal with average period ${\rm E}[T]$ for each trace. From this, we get measurements of E[$T$] and Var[$T$] at each value of $\Omega$ without any fitting.

We compare the experimental results for E$[T]$ and Var[$T$] that come directly from the clock signals to the theoretical predictions in Eq. (\ref{eq:jump_stat1}), showing similar values, albeit not exact to those predicted by the theory in Fig. \ref{fig:jump_data}.
This is likely due to unaccounted-for systematic errors in the signal that are retained after the low-pass filtering.
Furthermore, the comparison between theory and experiment for $N_{\rm jump}$ in the bottom panel of Fig. \ref{fig:Jumpinequality} (discussed further below) shows excellent agreement.

\section{\label{sec:thermodynamic_clocks}Uncertainty relation of a Quantum Clock}

So far, we have detailed theoretically and experimentally how our superconducting circuit can be operated as both a quantum oscillatory and quantum jump clock. 
We now connect these results with the (kinetic) thermodynamics of precision via an uncertainty relation. 
Specifically, we show that the precision of the clock ($N$) can be bounded by an uncertainty relation for the FPT, which is the period, $T$, of the clock.
Here we follow the derivation for the FPT uncertainty relation given in Refs.~\cite{Vu_thermodynamics_2022, Hasegawa_Quantum_2020}, which begins with the Cramer-Rao bound for the FPT:
\begin{equation}
\label{eq:cramer-Rao}
    \frac{{\rm Var}[T]}{\left(\partial_{\theta}{\rm E}_{\theta}[T]\vert_{\theta=0}\right)^{2}}\geq \frac{1}{I_{Q}(0)}\,.
\end{equation}
Here $\theta$ is an estimated parameter that we set to $0$, and ${I_{Q}(0)=I_{Q}(\theta=0)}$ is the quantum Fisher information (QFI) of the estimated parameter. 
We have used the fact that the QFI upper bounds the classical Fisher information, when maximized over all positive operator-valued measures: $I_{C}(0) \leq I_{Q}(0)$.
The choice of $\theta$ determines which physical quantity is bounded in the uncertainty relation. 
To obtain a bound for the FPT of the qubit dynamics given by Eq.~(\ref{eq:qubit_me}), the chosen parameterization is \cite{Vu_thermodynamics_2022}
\begin{equation}
\label{eq:parameterisation}
    \Omega \rightarrow (1+\theta)\Omega\,, \quad 
    \Delta \rightarrow (1+\theta)\Delta\,, \quad 
    \Gamma \rightarrow (1+\theta)\Gamma\,, \quad 
\end{equation}
which corresponds to speeding up the dynamics when $\theta>0$, and a slowing of the dynamics when $\theta < 0$.
Under such a parameterization, one can derive an expression for the QFI \cite{Gammelmark_2014}, which can be written in terms of the quantities \cite{Vu_thermodynamics_2022}
\begin{equation}
\label{eq:Qfisher}
    I_{Q}(0) = \mathrm{E}[T](\mathcal{N} + \mathcal{Q})\,,
\end{equation}
where 
\begin{equation}
    \mathcal{N} = \Gamma\,,\qquad \mathcal{Q} = \frac{4 \left[\Gamma ^2 \Delta ^2+\left(\Delta ^2+\Omega ^2\right)^2\right]}{\Gamma  \Omega ^2}\,,
\end{equation}
the derivation of which can be found in Appendix~\ref{sec:Uncertainty_derivation}.
In this expression, $\mathcal{N}$ is called the dynamical activity, which is a measure of the kinetic activity in the system and arises in KURs \cite{Di_Terlizzi_2018}.
It is worth noting, too, that the QFI is proportional to the inverse of the resolution of the clock, which is defined $R = 1/{\rm E}[T] $. 

Most importantly, we are now in a position to evaluate the left-hand side of Eq.~(\ref{eq:cramer-Rao}) for both the oscillatory and jump regimes. 
We evaluate $\partial_{\theta}{\rm E}[T]\vert_{\theta=0}$ using the respective distributions for the FPT given in Eq.~(\ref{eq:Wald}) and Eq.~(\ref{eq:Poisson}). 
In Appendix~\ref{sec:Uncertainty_derivation}, we show that in both regimes $\partial_{\theta}{\rm E}[T]\vert_{\theta=0} = - {\rm E}[T]$, thus allowing us to define the KUR valid in both regimes 
\begin{equation}
\label{eq:inequality}
    N \leq {\rm E}[T] (\mathcal{N} + \mathcal{Q})\,.
\end{equation}
We note that the derivation reveals that this expression is valid for any quantum clock whose timing signal's FPT obeys an inverse Gaussian distribution (though the expressions for $\mathcal{N}$ and $\mathcal{Q}$ are clock dependent).

\begin{figure}
    \centering
    \includegraphics[width=\columnwidth]{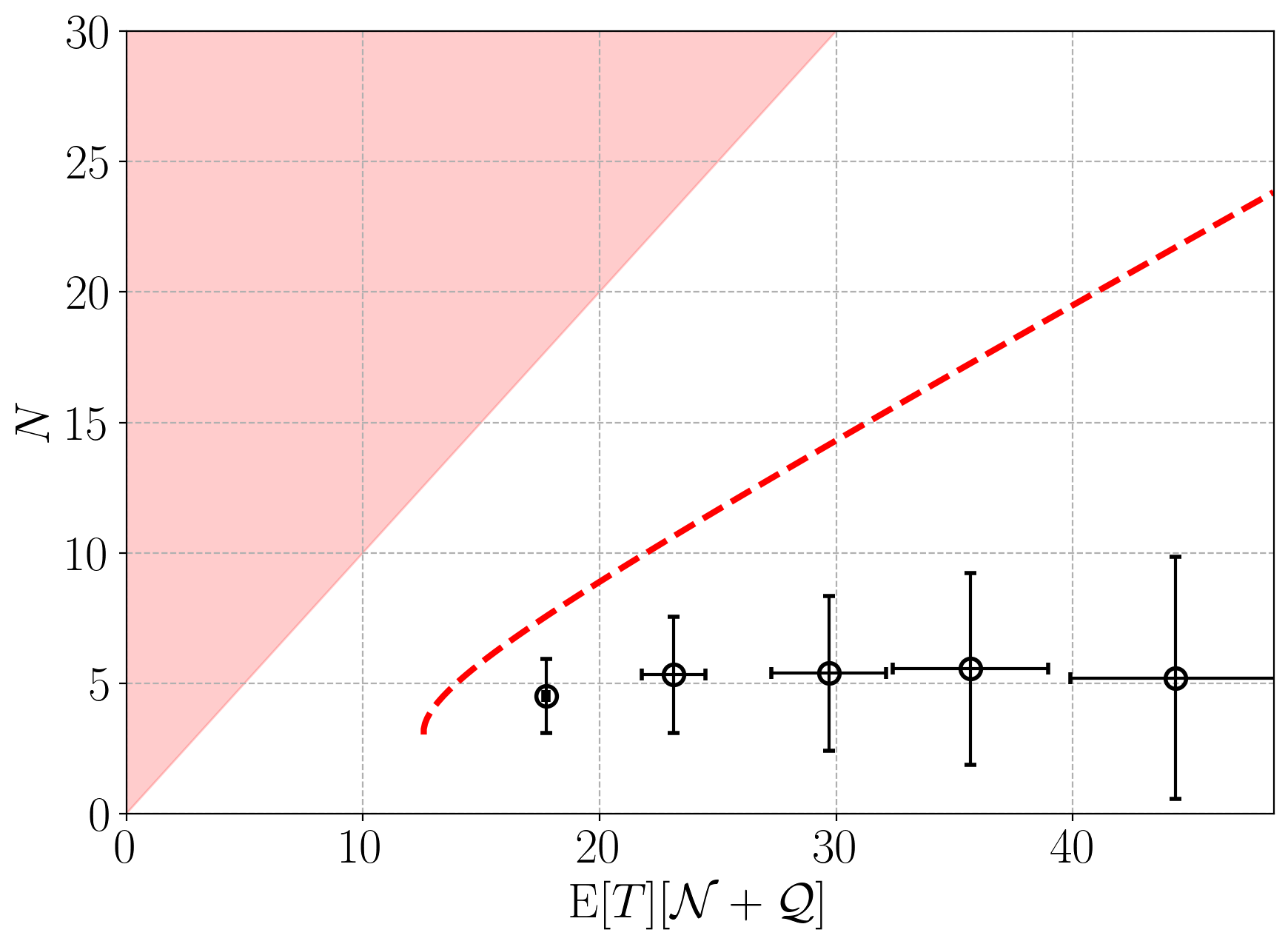}
    \caption{
    Precision ($N={\rm E}[T]^2$/${\rm Var}[T]$) versus the QFI. The red dashed line corresponds to the theoretical prediction $N=N_{\rm osc}$ [Eq.~(\ref{eq:N_osc})], and the shaded region signifies violation of the KUR [Eq. (\ref{eq:inequality})]. The black circles correspond to the experimental data at each of the five values of $\Omega$ given in Fig. \ref{fig:PSDs}. Both theory and experiment obey the KUR. The experimental $N$ does not increase with rising QFI (and, hence, rising $\Omega$) as much as theoretically predicted due to the large difference (seen in Fig.~\ref{fig:wald_data}) between the experimental and theoretical values of ${\rm Var}[T]$.}
    \label{fig:waldinequality}
\end{figure}

In Fig.~\ref{fig:waldinequality}, we compare the theoretical and experimental results regarding the QFI [Eq. (\ref{eq:Qfisher})] with the precision [$N_{\textrm{osc}}$, Eq.~(\ref{eq:N_osc})] and the experimental data at the five values of $\Omega$ given in Fig.~\ref{fig:PSDs}, at the mean cavity photon number $\vert \alpha_{0}\vert^{2} =0.28$ (the dephasing rate $\Gamma$ is determined as described in Appendix~\ref{sec:calibration}).
Both theory and experiment satisfy the KUR [Eq. (\ref{eq:inequality})] over the values of $\Omega$ that we test. However, for the larger values of $\Omega$---shorter tick times (${\rm E}[T]$) but larger $I_{Q}(0)$---the experimental $N_{\rm osc}$ increases at a slower rate than predicted by the theory, which is largely driven by the error in our estimate of ${\rm Var}[T]$ as seen in Fig.~(\ref{fig:wald_data}). 

\begin{figure}
    \centering
    \includegraphics[width=\columnwidth]{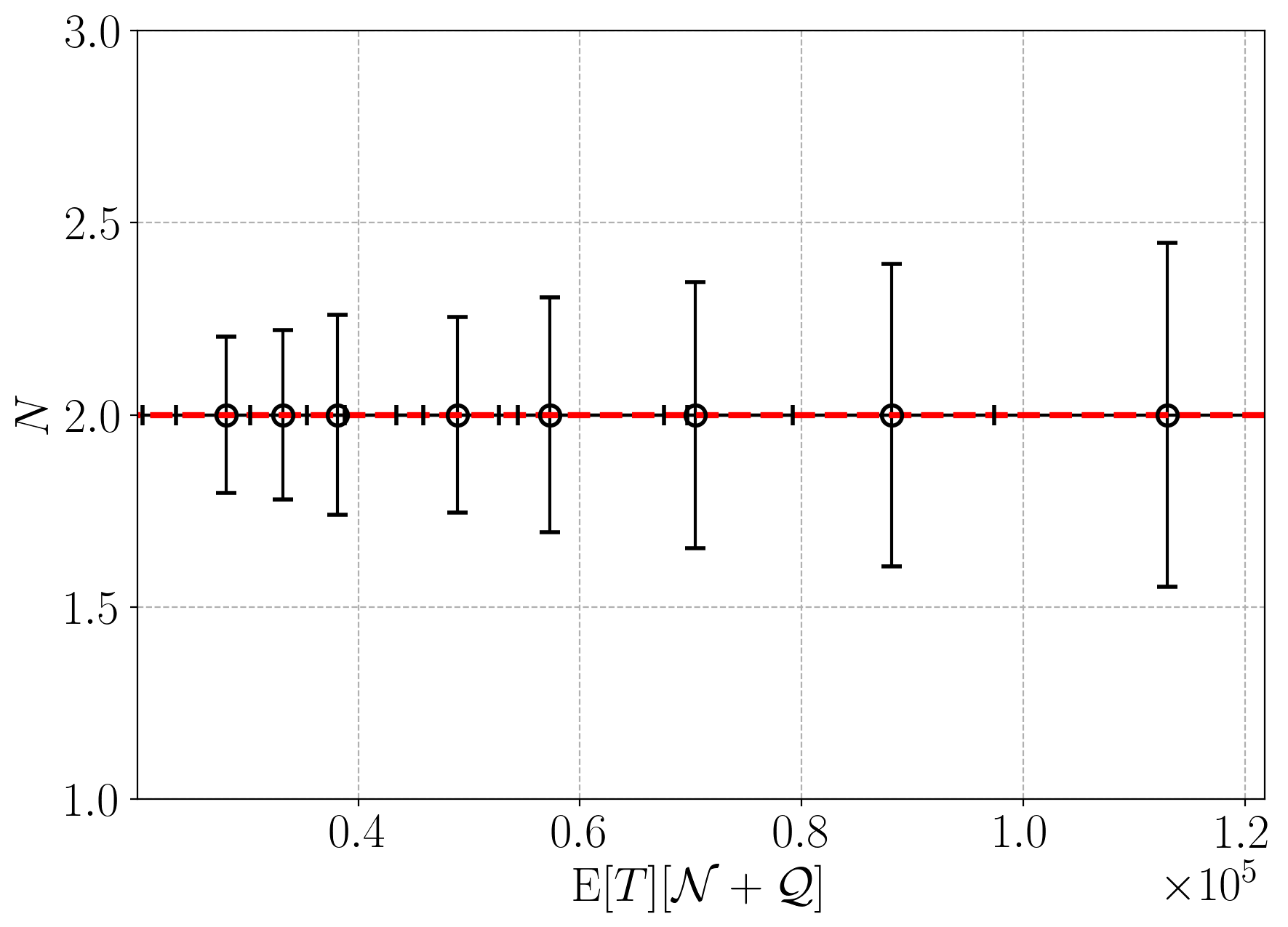}
    \caption{ Precision ($N$) versus the QFI. The red dashed line corresponds to the theoretical prediction $N=N_{\rm jump}$ [Eq.~(\ref{eq:jump_stat1})]. The black circles correspond to the data taken at different values of $\Omega$. In this domain, theory and experiment agree very well, and both obey the KUR [Eq.~(\ref{eq:inequality})]. Note that the region of KUR violation is not visible on this plot.
    }
    \label{fig:Jumpinequality}
\end{figure}

Using the calibrated parameters in the jump regime, we can also compute both the theoretical and experimental values of the QFI and $N$.
In Fig.~\ref{fig:Jumpinequality}, we plot $N$ as a function of the QFI that shows very good agreement with the theoretically predicted values. 
As with the oscillatory clock, the KUR is both theoretically and experimentally satisfied by the jump clock in the tested regime.
We also find excellent agreement between theory and experiment for both $N$ and $I_{Q}(0)$ at all chosen values of $\Omega$ in the regime $\Omega/2\pi>$243 kHz.
Given the large values of $I_{Q}(0)$, the Poisson distribution very comfortably satisfies the KUR.

\section{Summary and Discussion}
\label{sec:conclusion}

Fundamental limits of quantum clocks are of relevance to both fundamental physics and technological development. All clocks that generate classical timing signals  necessarily entail measurement, which for quantum clocks impacts their precision due to the collapse postulate of quantum mechanics.  Here we have shed light on how the measurement backaction fundamentally affects the precision of quantum clocks by studying the specific case of a coherently driven quantum clock that is continuously measured and exhibits dynamics of two types: quantum coherent (i.e. oscillatory clock) and incoherent (i.e. jump clock).

On the theoretical side, we used a KUR for the period (i.e. first passage time) of the clock's timing signal, which set a fundamental bound on the precision of the clock in terms of the dynamical activity---a measure of the system's kinetic dynamics---which is more broadly connected to the quantum Fisher information. The KUR contained an explicit dependence on the measurement strength. Independent of the KUR, from a quantum optical point of view, we derived equalities for the precision of the clock for both the coherent and incoherent regimes, and found that these equalities obeyed the KUR.  The equality for the precision was a constant in the incoherent regime, but the equality in the quantum coherent regime was inversely proportional to the measurement strength and directly proportional to the clock drive strength. This revealed that the quantum coherent regime of the clock, in principle, could attain the unbounded growth in precision permitted by the KUR.

On the experimental side, we measured the precision of the clock in both the coherent and incoherent regimes, and always found the KUR to be satisfied.  This constituted an experimental test of a kinetic uncertainty relation for any quantum system.  
In the quantum jump regime, we found excellent quantitative agreement between the theory and experiment for the clock's precision, despite the slight discrepancy in the measured tick times and their variance.
In the quantum coherent regime, the experimental results qualitatively behaved according to the theory, but due to systematic noise that could not be accounted for, we did not get strict quantitative agreement between the theory and experiment.  
It remains an open problem for future work to obtain experimental confirmation of the theoretically predicted linear growth of the precision in the quantum coherent regime since it would demonstrate that quantum clock precision can in principle be enhanced even in situations where quantum measurement backaction degrades it.

It is interesting to note that for the incoherent regime of our quantum clock, we theoretically and experimentally found that at sufficiently large drive strengths the precision was saturated, while for the incoherent quantum clock model in Ref. \cite{Erker}, they theoretically found that at sufficiently large drive strengths the \textit{accuracy} was saturated. \footnote{Ref. \cite{Erker} actually reported saturation of the accuracy with increasing fundamental heat dissipation of the drive instead of increasing drive strength, but the latter is a monotonically increasing function of the former.} In contrast, the quantum coherent regime of our clock theoretically showed a precision with no upper bound---ignoring practical considerations.

Regarding the possibility of technological applications, the particular clock that we studied here is not fully amenable to microscale integration due to the requirements of a coherent drive and homodyne measurement. However, we note that recent theoretical work \cite{Manikandan} has shown that the coherent drive could, in principle, be replaced by a suitably engineered coupling to the environment.  Further consideration of how to engineer a full microscale quantum coherent clock inclusive of the measurement apparatus remains an open problem.

Finally, in this work, we have demonstrated that current superconducting circuit technology is capable of performing tests of KURs in the quantum domain.  This encourages consideration of how this platform may be used to perform further such tests.  While the KURs in such tests are necessarily particular to the system at hand, the underlying theoretical ideas that are used for their derivation are more general in applicability, and each test of a particular KUR provides further validation of those more general ideas.  It is for this reason that experimental tests of KURs are of broad fundamental importance.

\section*{Acknowledgements}
This research project was supported by the Foundational Questions Institute Fund, a donor-advised fund of Silicon Valley Community Foundation, under Grant No. FQXi-IAF19-04. We also acknowledge support from the Australian Research Council Centre of Excellence for Engineered Quantum Systems (EQUS, CE170100009). M.J.K. acknowledges financial support from a Marie Sk\l odwoska-Curie Fellowship (Grant No. 101065974).


\newcommand{\beginsupplement}{
        \setcounter{table}{0}
        \renewcommand{\thetable}{S\arabic{table}}
        \setcounter{figure}{0}
        \setcounter{equation}{0}
        \setcounter{page}{1}
        \renewcommand{\thefigure}{S\arabic{figure}}
        \renewcommand{\thesection}{S\arabic{section}}
        }

\twocolumngrid

\appendix
\addcontentsline{toc}{section}{Appendix} 
\part{Appendix} 
\parttoc 

\section{\label{sec:theory} Theory}

\subsection{\label{sec:adiabatic_elimination}Adiabatic Elimination and Energy Dissipation}
A simplified description of Eq. (\ref{eqn:EOM_homodyne_main}) based on the adiabatic elimination of the field gives insight into the mechanism of the clock and its thermodynamic properties. In the absence of the dispersive coupling, the steady-state field inside the driven cavity is given by a coherent state with amplitude 
\begin{equation}
    \alpha_0=-\frac{2iE}{\kappa}
\end{equation}
It is thus useful to move to a displacement picture using a canonical transformation 
$\rho_c\rightarrow \rho'_c=D(\alpha_0)\rho_cD^\dagger(\alpha_0)$, where $D(\alpha)$ is the usual displacement operator. In this picture, the field remains close to the vacuum state, and we may use the approximation
\begin{align}
    \rho'_c(t) \approx& ~\rho_{00}\otimes|0\rangle\langle 0|+\rho_{01}\otimes|0\rangle\langle 1|\nonumber\\
    &+\rho_{10}\otimes|1\rangle\langle 0|+\rho_{11}\otimes|1\rangle\langle 1|,
\end{align}
where the $\rho_{jk}$ are operators that act only on the atom Hilbert space. The reduced density for the qubit is $\rho_A=\rho_{00}+\rho_{11}$. 
The conditional master equation in the displacement picture is given approximately by 
\begin{align}
   d\rho_A =&   - i\Omega[\sigma_x,\rho_A]dt -i\Delta[\sigma_z,\rho_A]dt\nonumber\\ 
   & +\gamma{\cal D}[\sigma_-]\rho_A dt+\Gamma{\cal D}[\sigma_z]\rho_A dt \nonumber\\
   &-\sqrt{\Gamma}{\cal H}[\sigma_z]\rho_A dW(t),
   \label{displace_homodyne}
\end{align}
where $g=\chi|\alpha_0|$,  $\Delta = \chi|\alpha_0|^2$, and $\Gamma=4g^2/\kappa$. The term proportional to $\Delta$ describes the detuning of the qubit due to the average photon number in the uncoupled cavity; that is, $\Delta$ is the effective Stark shift of the qubit due to the dispersive interaction with the cavity field. Note that, as both $\Gamma$ and $\Delta$ are linear in the mean photon number in the cavity, $\Gamma=4\chi\Delta$. The rate of decay of the off-diagonal matrix elements, in the eigenstates of $\sigma_z$, is $2\Gamma$.  

The average field emitted by the cavity is given by
\begin{equation}
    \langle a_{out}(t)\rangle=-\frac{g}{\sqrt{\kappa}}\langle \sigma_z(t)\rangle \, .
\end{equation}
This has units such that the average intensity  $\langle a_{out}^\dagger(t)a_{out}(t)\rangle$ has units of photons per second. The measured homodyne photocurrent at the output field is given, in the long-time limit, by
\begin{eqnarray}
\label{eqn:Rabi_current_homodyne}
J_x(t)dt & = & g\sqrt{\eta}\langle \sigma_z\rangle_c dt+ \sqrt{\kappa}dW(t),
\end{eqnarray}
where the conditional average is computed using $\rho_c'(t)$ and $\eta$ is the efficiency of the homodyne detection. 

Expanding the atomic density operator in the Pauli basis, the master equation is equivalent to the stochastic differential equations
\begin{eqnarray}
dX & = & -2\Delta Y dt -\gamma_2 X dt \label{x-dynamics}\\
dY & = & 2\Delta X dt -2\Omega Z dt -\gamma_2 Y dt \label{y-dynamics}\\
dZ & = & 2\Omega Y dt-\gamma(1+Z) dt-2\sqrt{\Gamma}(1-Z^2)dW\label{z-dynamics}
\end{eqnarray}
where $\gamma_2 = \gamma/2+2\Gamma$ is the transverse decay rate of the conditional polarization.  The spontaneous emission rate in our experiment is very small,  and we ignore it so that $\gamma_2= 2\Gamma$.

The unconditional dynamics is obtained by averaging over the noise. This means that the equations of motion for the unconditional state are found by setting the last term in Eq. (\ref{z-dynamics}) to zero. If we change variables to $\bar{X},\bar{Y},\bar{Z}$ by a rotation around the $Y$ axis of angle $\theta$ such that $\cos(\theta)=\Omega/\tilde{\Omega}$, where $\tilde{\Omega}=\sqrt{\Omega^2+\Delta^2}$, we find that conservative dynamics describe precession around the $\bar{Z}$ axis at rate $2\tilde{\Omega}$. This is the effective Rabi frequency in the model.  The inclusion of damping modifies this frequency slightly. The exact value is given by the imaginary part of the eigenvalues of the dynamics for $X,Y,Z$. The unconditional dynamics exhibits either underdamped dynamics (damped oscillations) or overdamped dynamics. This corresponds to a change of the eigenvalues from complex to real as $\Gamma$ is varied. In the case of $\Delta=0$, complex roots occur for $\Omega >\Gamma/2$.
In the conditional dynamics, we expect sustained, but noisy, oscillatory dynamics to correspond to the underdamped regime in the unconditional dynamics.

As both $\Delta$ and $\Gamma$ depend on $n_0=|\alpha_0|^2$, the mean photon number in the cavity, it is of interest to determine the dependence of the effective Rabi frequency, $2\tilde{\Omega}$, on $n_0$. In units of the cavity linewidth, we see that
\begin{eqnarray}
\frac{\Gamma}{\kappa} & = & 4\left (\frac{\chi}{\kappa}\right )^2 n_0\\
\frac{\Delta}{\kappa} & = & \left (\frac{\chi}{\kappa}\right ) n_0
\end{eqnarray}

The full cavity-qubit dynamics using a full numerical simulation of the complete system with experimentally realistic parameters is shown in Fig. 9 . We numerically integrate Eq. (\ref{eqn:EOM_homodyne_main}) and apply low-pass filtering to resolve the clock signal.  The maximum cavity occupation in the simulations is set at ten photons. Computational constraints make going much higher than this difficult as the size of the Hilbert space becomes too large.    The simulations are performed with QuTiP \cite{johansson2012qutip,johansson2013qutip}.

\begin{figure}
\centering
\includegraphics[width=\columnwidth]{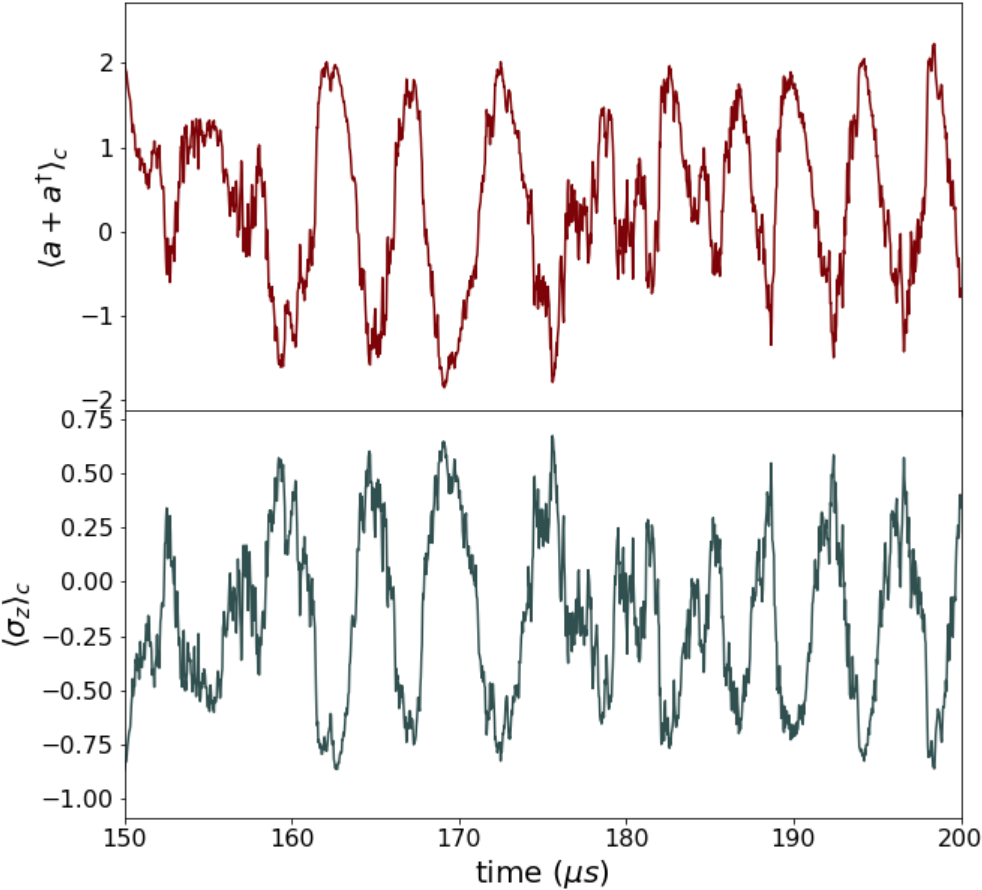}
\includegraphics[width=\columnwidth]{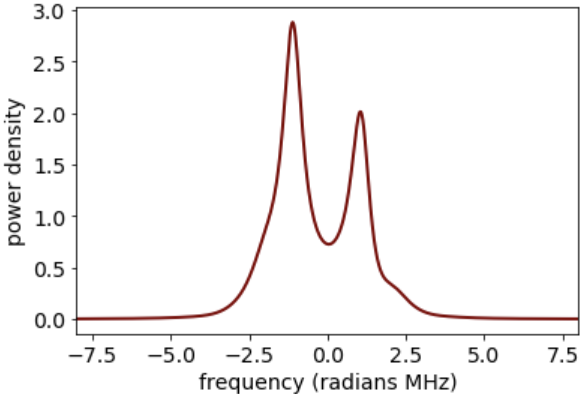}
\caption{(Top) When the cavity output of the superconducting device is continuously measured with the scheme of Fig. \ref{main:homodyne-scheme}, as simulated with Eq.~(\ref{eqn:EOM_homodyne_main}), the qubit exhibits noisy oscillations in its conditional long-time dynamics for a cavity drive strength of $E/2\pi=0.25$. Here the system parameters are (in megahertz) given by $\chi=2$, $\gamma=0.2$, $\kappa=1$, $\Omega=1$, and $\eta=0.2$. The expectation value of the monitored cavity quadrature, $\langle a + a^{\dagger}\rangle_c(t)$, closely tracks the qubit population dynamics, $\langle \sigma_z\rangle_c(t)$.
(Bottom) The cavity emission power spectrum is obtained from the top figure. The positive frequencies of the power spectrum are a measure of the oscillator's capacity to absorb energy, while the negative frequency part is a measure of its capacity to emit energy; the asymmetry in the spectrum merely indicates the driven nature of the cavity. 
The cavity emission power spectrum clearly shows the Rabi oscillations of the qubit imprinted on the cavity field in the long-time limit.}
\label{fig_Rabipowerspectrum_homodyne}
\end{figure}

We obtain the Rabi oscillation clock regime by setting $E/2\pi =  0.25$. We depict these simulations in the top panel of Fig. \ref{fig_Rabipowerspectrum_homodyne}, which shows that the resulting conditional dynamics of the qubit population [$\langle \sigma_z \rangle_c(t)$] has noisy oscillations, and the cavity field $\hat{x}$ quadrature [$\langle a + a^{\dagger}\rangle_c(t)$] tracks the evolution of $\langle\sigma_z\rangle_c(t)$. 
The cavity emission power spectrum [$S(\omega)=\int \langle a^{\dagger}(\tau)a(0)\rangle e^{-i\omega\tau}d\tau$] shown in the lower panel of Fig. \ref{fig_Rabipowerspectrum_homodyne} verifies that the Rabi oscillations do not decay with time.

\begin{figure}
\centering
\includegraphics[width=\columnwidth]{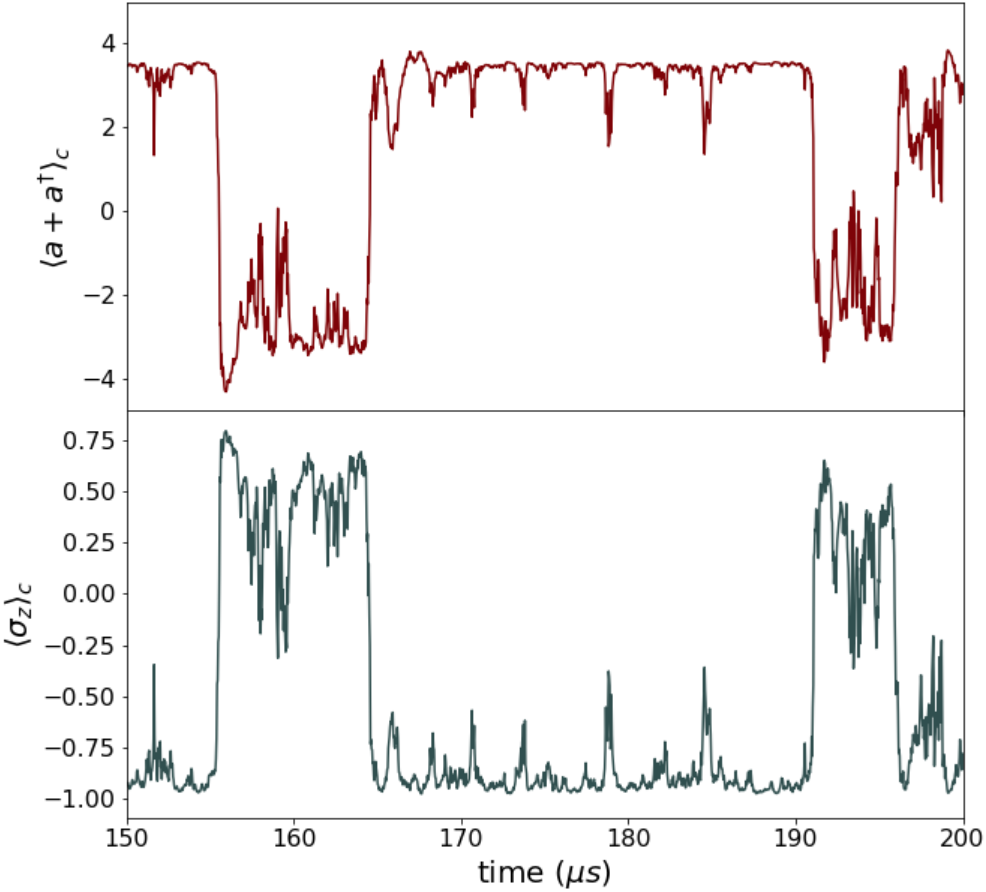}
\includegraphics[width=\columnwidth]{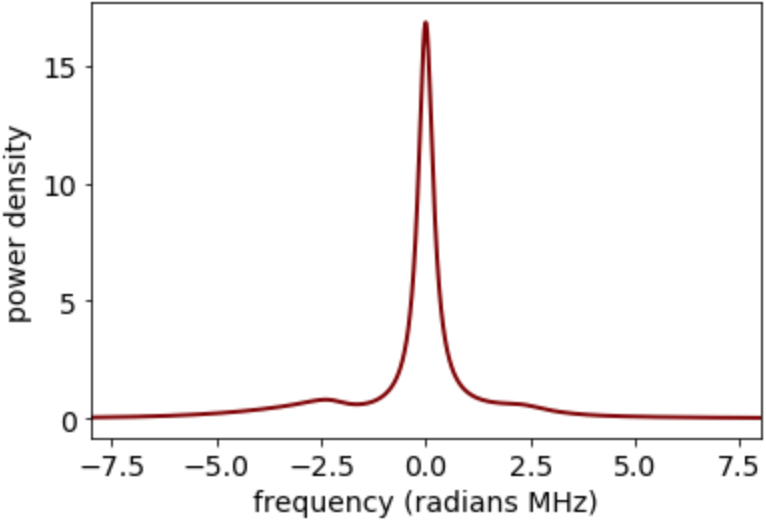}
\caption{(Top) When the cavity drive is sufficiently large (here $E/2\pi=0.6$), the measurement on the qubit becomes strong enough to give rise to the quantum Zeno effect, and the qubit exhibits quantum jumps between its ground and excited states. Here the system parameters are in megahertz and given by $\chi=2$, $\gamma=0.2$, $\kappa=1$, $\Omega=1$, and $\eta=0.2$. 
(Bottom)
Here the peak at zero indicates a high degree of correlation between the system's current state and its near future state; given the Zeno dynamics, this makes sense as the dynamics are largely frozen except for when there is a sudden jump to the other state. 
In contrast to the case of Rabi oscillations (see Fig. (\ref{fig_Rabipowerspectrum_homodyne})), in the quantum jump regime, the spectral weight of the cavity field output is peaked around zero. The positive frequencies are a measure of the oscillator's capacity to absorb energy, while the negative frequencies are a measure of its capacity to emit energy; the asymmetry in the spectrum is due to the driven nature of the cavity.}
\label{fig_qjumpspowerspectrum_homodyne}
\end{figure}

We can enter the quantum jump regime by increasing the cavity drive strength to $E/2\pi=0.6$ while keeping the other parameters the same.  An illustration is shown in the top panel of Fig. \ref{fig_qjumpspowerspectrum_homodyne}.
The corresponding cavity emission power spectrum $S(\omega)=\int \langle a^{\dagger}(\tau)a(0)\rangle e^{-i\omega\tau}d\tau$ shown in Fig. \ref{fig_qjumpspowerspectrum_homodyne} reveals that the spectral weight is peaked around zero, confirming the suppression of Rabi oscillations.

\subsection{First Passage Time in the Oscillatory Dynamics}
\label{sec:Oscillatory Dynamics}

We first consider the underdamped case, for which noisy oscillations occur in the measured current. In our experiment, the spontaneous emission rate for the qubit is very low: $\gamma/\kappa=0.000\,56$.  Therefore, once $n_0\neq 0$, the irreversible dynamics of the qubit is dominated by the dephasing arising from the dispersive measurement. Using our parameters, the condition for underdamped motion is $\Omega > \Gamma/2$ (corresponding to complex eigenvalues in the unconditional dynamics when $\Delta=0$). In terms of the mean photon number in the cavity, this implies that $\Omega > 2(\chi/\kappa)^2\kappa n_0$. In terms of the driving field, $\Omega > 8\chi^2|E_0|^2/\kappa^3$.  We call this the weak measurement limit. Loosely speaking, we can say that in this limit, the driving power responsible for quantum coherence exceeds the energy dissipated due to the measurement channel.  

If we neglect spontaneous emission in the stochastic differential equations, Eqs. (\ref{x-dynamics}-\ref{z-dynamics}), the conditional state remains on the Bloch sphere.  We thus change to spherical polar coordinates for the unit sphere. The stochastic differential equations then become
\begin{eqnarray}
d\phi & = & 2\Delta dt-2\Omega \cot\theta \cos\phi dt\\
d\theta & = & -2\Omega\sin\phi dt +2\sqrt{\Gamma}\sin\theta dW
\end{eqnarray}
The noise is multiplicative, as it must be: it turns off when $Z^2=1$. However, we can approximate this with linear noise as follows.   

In the case that $\Delta << \Omega$, the orbit is mostly confined to the $Y$-$Z$ plane (precession around the $X$ axis), so we set $\phi=\pi/2$, and then to a good approximation
\begin{equation}
 d\theta =  -2\Omega dt +2\sqrt{\Gamma}\sin\theta dW.
\end{equation}
In the weak measurement limit defined above we have small $\sqrt{\Gamma}/\Omega$, and the noise can therefore be linearized by replacing $\sin\theta$ by its rms value over one cycle so that 
\begin{equation}
\label{phase-diffusion}
 d\theta =  -2\Omega dt +\sigma  dW,
\end{equation}
where $\sigma=\sqrt{2\Gamma}$. A similar equation describes the phase diffusion on a classical limit cycle used in conventional clocks.  However, in that case, $\sigma$ is proportional to temperature \cite{Milburn-CP}, while here the phase diffusion is entirely due to quantum measurement noise.  

The statistics of the period can now be determined by the  distribution of the first passage time to move through an angle of $2\pi$. This distribution is the inverse Gaussian distribution (also known as the Wald distribution), given by \cite{Aminzare} 
\begin{equation}\label{eq:wald2}
    P(T)= \sqrt{\frac{2\pi}{\sigma^2 T^3}}\exp\left [-\frac{(2\pi-2\Omega T)^2}{2\sigma^2 T}\right ].
\end{equation}

\subsection{First Passage Time in the Jump Dynamics}
\label{sec:Jump Dynamics}

We now turn to the quantum jump regime, where the clock signal is an aperiodic Markov jump process.  To see this, we note that this requires a fast measurement rate ($\Gamma$) such that the unconditional dynamics is underdamped. Thus we require $2\Omega < \Gamma$, or in terms of the cavity driving field, $\Omega < 8\chi^2|E_0|^2/\kappa^3$. In this case, the measurement rate $\Gamma$ is large, and the qubit is rapidly dephased. Loosely speaking, we can say that in this limit, the energy dissipated due to the measurement channel exceeds the driving power responsible for quantum coherence.   We can then approximate the master equation in Eq. (\ref{displace_homodyne}) with a classical Markov stochastic process for the occupation probabilities. This is given by 
\begin{equation}
\label{jump-me}
    \frac{d p_1}{dt}= -\nu p_1+\mu p_0,
\end{equation}
where 
\begin{eqnarray}
\mu & = & \frac{2\Omega^2}{\gamma+4\Delta^2/\gamma},\\
\nu & = & \gamma +\mu.
\end{eqnarray}
This is a discrete homogeneous Markov process with transition $0\rightarrow 1$ occurring at rate $\mu$, and the transition $1\rightarrow 0$ occurring at rate $\nu$.  The stochastic differential for the stochastic mean of $\sigma_z$ is thus
\begin{equation}
    dZ(t)= (1-Z)dN_+(t)-(1+Z)dN_-(t),
\end{equation}
where $dN_{\pm}(t)$ are Poisson point processes such that ${\cal E}[dN_+(t)]=\mu dt$ and ${\cal E}[dN_-(t)]=\nu dt$. 

This is like a quantum version of a Mach thermal clock \cite{MilMil}. 
The period of the clock ($T$) is defined as the time taken for two transitions to bring the clock back to the same state. The probability distribution is given by 
\begin{equation}
    P(T)= \frac{\mu\nu}{\nu-\mu} (e^{-\mu T}-e^{-\nu T})
\end{equation}
It is the same regardless of which state, $Z(0)=\pm 1$, is taken as the start of the period.  In the limit $\nu\rightarrow \mu$, this becomes
\begin{equation}
   P(T)=\mu^2 Te^{-\mu T}.
\end{equation}
The mean and variance of this distribution are
\begin{equation}
\label{eq:jump_stat2}
   {\rm E}[T] = \frac{2}{\mu}, \quad {\rm Var}[T]=\frac{2}{\mu^{2}}.
\end{equation}

\subsection{\label{sec:Uncertainty_derivation}Uncertainty relation for the FPT}

At it's core, all of the bounds stem from the quantum Fisher information and the Cramer-Rao inequality.
Now we introduce some parameters ($\theta$) that allow us to associate changes in the system dynamics with changes in $\theta$. 
Here $\theta$ is a list of numbers that could be used to indicate when jumps or other dynamical events occur. 
To derive an inequality for the FPT, the Liouvillian is parameterized according to \cite{Vu_thermodynamics_2022}
\begin{equation}
    H \rightarrow (1+\theta)H\quad L\rightarrow \sqrt{1+\theta}L\,,
\end{equation}
where $H = \Omega \sigma_{x}$ and $L = \sqrt{\Gamma}\sigma_{z}$, and we have assumed $\gamma$ to be negligibly small.
This parameterization corresponds to quickening and slowing of the system dynamics and can be viewed as scaling all the parameters according to Eq.~(\ref{eq:parameterisation}).
Importantly, this allows us to parameterize the Liouvillian in terms of
\begin{equation}
    \mathcal{L} \rightarrow \mathcal{L}_{\theta_{L}, \theta_{R}}\,, 
\end{equation}
where we introduce parameters for the left and right eigenstates of $\mathcal{L}$.
Here, for simplicity, we assume that there is only a single $\theta=\theta_{1}=\theta_{2}$ for both the left and right eigenstates, but the result in Ref. \cite{Gammelmark_2014} holds for a general number.
As shown in Ref.~\cite{Gammelmark_2014}, the QFI can written as
\begin{equation}
\label{eq:gammel}
    I_{Q}(0) = 4 T {\rm Re}\{\partial_{\theta_{L}}\partial_{\theta_{R}}\lambda(\theta_{L},\theta_{R})\}\vert_{\theta_{L} = \theta_{R}=0}\,,
\end{equation}
where $\lambda(\theta_{L},\theta_{R})$ corresponds to the dominant eigenvalue that coincides with the eigenvalue of the steady state $\rho_{\rm ss}$ in the unperturbed dynamics.
We can now substitute this back into the Cramer-Rao bound and evaluate it.
Then, the QFI can be evaluated over an average single cycle $T={\rm E}[\tau]$ \cite{Gammelmark_2014, Vu_thermodynamics_2022}
\begin{equation}
    I_{Q}(0) = E[\tau] \left(\mathcal{N} + \mathcal{Q}\right)\,,
\end{equation}
where both of these terms can be evaluated using the steady state of $\mathcal{L}$: $\rho_{\rm ss} = \mathbb{I}/2$.  This yields
\begin{equation}
    \mathcal{N} = \tr{L^{\dagger}L\rho_{ss}} = \Gamma\,,
\end{equation}
which is called the dynamical activity of the unconditional master equation \cite{Vu_thermodynamics_2022}.
The second term, written in vectorized notation, is 
\begin{align}
    \mathcal{Q} &= -4\left( \bra{I} \mathcal{K}_{1}\mathcal{L}^{+}\mathcal{K}_{2}\ket{\rho_{ss}} + \bra{I} \mathcal{K}_{2}\mathcal{L}^{+}\mathcal{K}_{1}\ket{\rho_{ss}}\right) \\
    &=\frac{4 \left[\Gamma ^2 \Delta ^2+\left(\Delta ^2+\Omega ^2\right)^2\right]}{\Gamma  \Omega ^2},
\end{align}
where $\mathcal{K}_{1}\rho = - i H \rho + (L^{\dagger}\rho L - L^{\dagger}L \rho)/2$ and $\mathcal{K}_{2}\rho = i\rho H + (L^{\dagger}\rho L - \rho L^{\dagger}L )/2$ are two superoperators and $\mathcal{L}^{+}$ is the Drazin inverse, which is directly related to the Moore-Penrose inverse. 
Bringing these together and doing a little rearranging, we can now define the right-hand side of the inequality:
\begin{equation}
\label{eq:KUR}
    \frac{{\rm Var}[\tau] }{(\partial_{\theta}{\rm E}_{\theta}[\tau])^{2}\vert_{\theta=0}}\, \geq \frac{1}{{\rm E}\left[\tau\right]\left(\mathcal{N} + \mathcal{Q}\right)}\,.
\end{equation}

We want to study this inequality for both the oscillator and jump regimes, but we begin with the oscillator regime ($\Gamma < 2 \Omega$).\
Now we want to compute $\partial_{\theta}{\rm E}_{\theta}[\tau]$ for the inverse Gaussian (Eq.~(\ref{eq:Wald})).
Using the parameterization in Eq.~(\ref{eq:parameterisation}) this yields
\begin{align}
    E_{\theta}[T] 
    &=\frac{1}{1+\theta}\int_{0}^{\infty}dT' T' \sqrt{\frac{\pi}{\Gamma T'^{3}}}\exp \left(-\frac{(\pi - \Omega T' )^{2}}{\Gamma T'}\right)\,,
\end{align}
where we have introduced the change of variable $T' = (1+\theta)T$. Thus, we can easily show that 
\begin{equation} \label{eq:dE}
    \partial_{\theta}{\rm E}_{\theta}[T]\vert_{\theta=0} = - E[T]\,,
\end{equation}
which can now be substituted back into the Cramer-Rao bound (Eq. (\ref{eq:cramer-Rao})) to yield Eq.~(\ref{eq:inequality}). We note that Eq. (\ref{eq:dE}) is valid for any inverse Gaussian distribution of $T$, not only that given in Eq. (\ref{eq:Wald}), and therefore applies to any quantum clock where $T$ obeys an inverse Gaussian distribution.

In the jump regime ($\Gamma > 2\Omega$), we know that the dynamics is much better approximated by a Poisson process and is given by Eq.~(\ref{eq:Poisson}).
Now we want to compute $\partial_{\theta}{\rm E}_{\theta}[\tau]$ using the transformation in Eq.~(\ref{eq:parameterisation}), which yields 
\begin{align}
    E_{\theta}[\tau] =\frac{1}{1+\theta}\int_{0}^{\infty}T'\mu^{2} T' e^{-\mu T'}\,
\end{align}
where we have again used the transformation $T' = (1+\theta)T$.  This again yields
\begin{equation}
    \partial_{\theta}{\rm E}_{\theta}[T]\vert_{\theta=0} = - E[T]\,,
\end{equation}
which can now be substituted back into the Cramer-Rao bound to yield Eq.~(\ref{eq:inequality}).

\section{Experimental details}
\label{section_experimental_details}

\subsection{\label{sec:qubit_tuning}Qubit tuning}
In the experiment, the transmon qubit frequency at the symmetry point (where qubit frequency is maximum and decoherence due to flux noise is minimal) is 6.1 $\mathrm{GHz}$, and the longitudinal ($T_1$) and transverse ($T_2$) relaxation times of the qubit are respectively $10$ and $15\,\upmu\mathrm{s}$. In order to increase the $T_1$ of the qubit, we tune the qubit frequency down to $4.75\,\mathrm{GHz}$. This reduces the dispersive shift ($\chi/2\pi$) from $2.5\,\mathrm{MHz}$ to $340\,\mathrm{kHz}$, and changes $T_1$ and $T_2$ to the values given in the main text.

\subsection{Qubit drive amplitude calibration}
\begin{figure}[ht!]
\centering
\includegraphics[width=1\columnwidth]{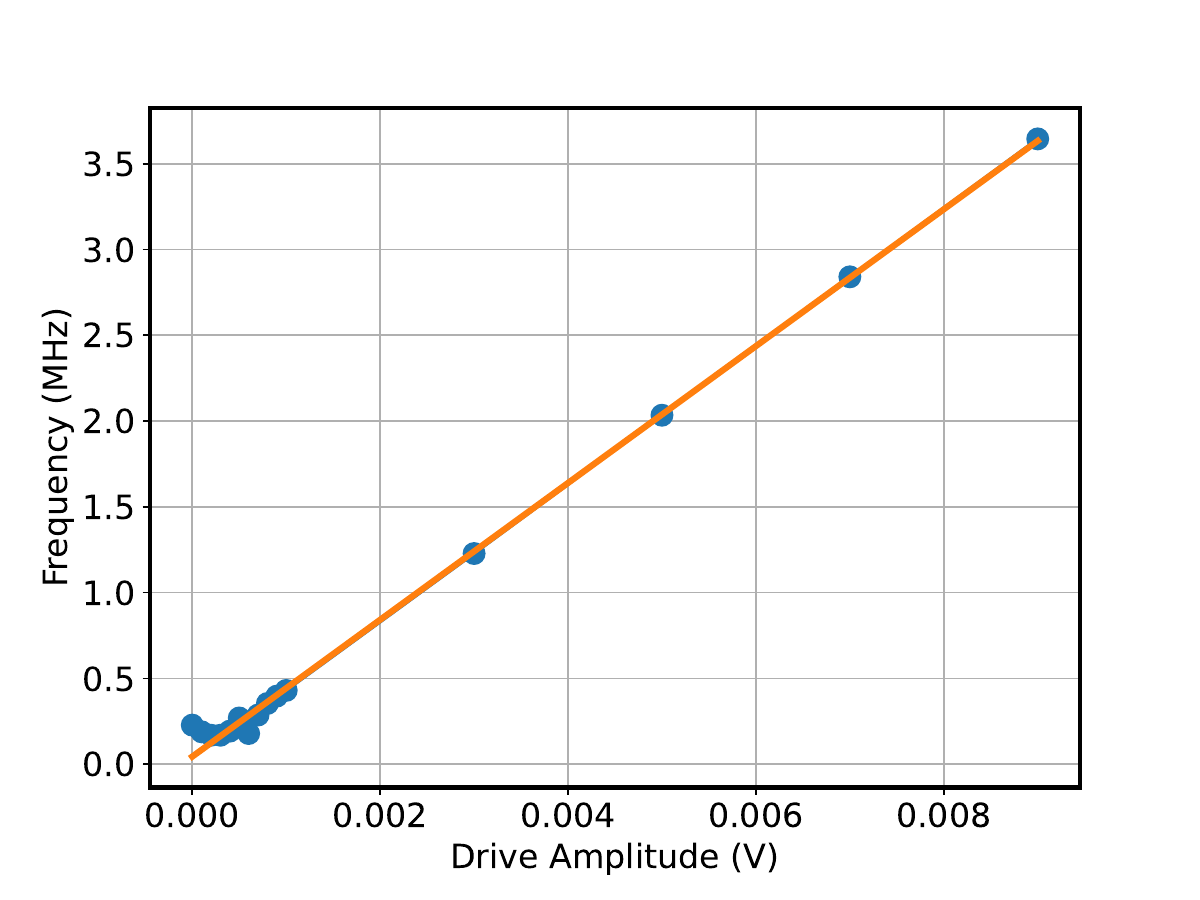}
\caption{The Rabi frequency extracted from fitting the Rabi oscillations in the time domain to each trace in a large ensemble and taking the ensemble average is linearly proportional to the drive amplitude, and the slope is the same as that found with the single-trace fitting in Fig.~\ref{fig:PSD_freqs}.}
\label{fig:2D_rabi}
\end{figure}

 Here we discuss how we calibrate the qubit drive amplitude to the Rabi oscillation frequency. At a given qubit drive amplitude, we perform a fitting to each time-domain trace of the Rabi oscillations in a large ensemble of traces.  Each fitting yields a Rabi frequency, and the ensemble average of these Rabi frequencies yields the value of the Rabi frequency that is assigned to the given qubit drive amplitude.  Fig. \ref{fig:2D_rabi} shows that this results in a linear relationship between the qubit drive amplitude and the assigned Rabi frequencies. An alternative calibration (see Appendix \ref{sec:calibration_driveamp_freqdomain}) at a different set of qubit drives and with fitting in the frequency domain yields the same slope (and zero offset). This confirms that the calibration of the qubit drive amplitude is reliable.
 
\subsection{Spectrum of the oscillations in the weak measurement regime}

\begin{figure}
\centering
\includegraphics[width=1\columnwidth]{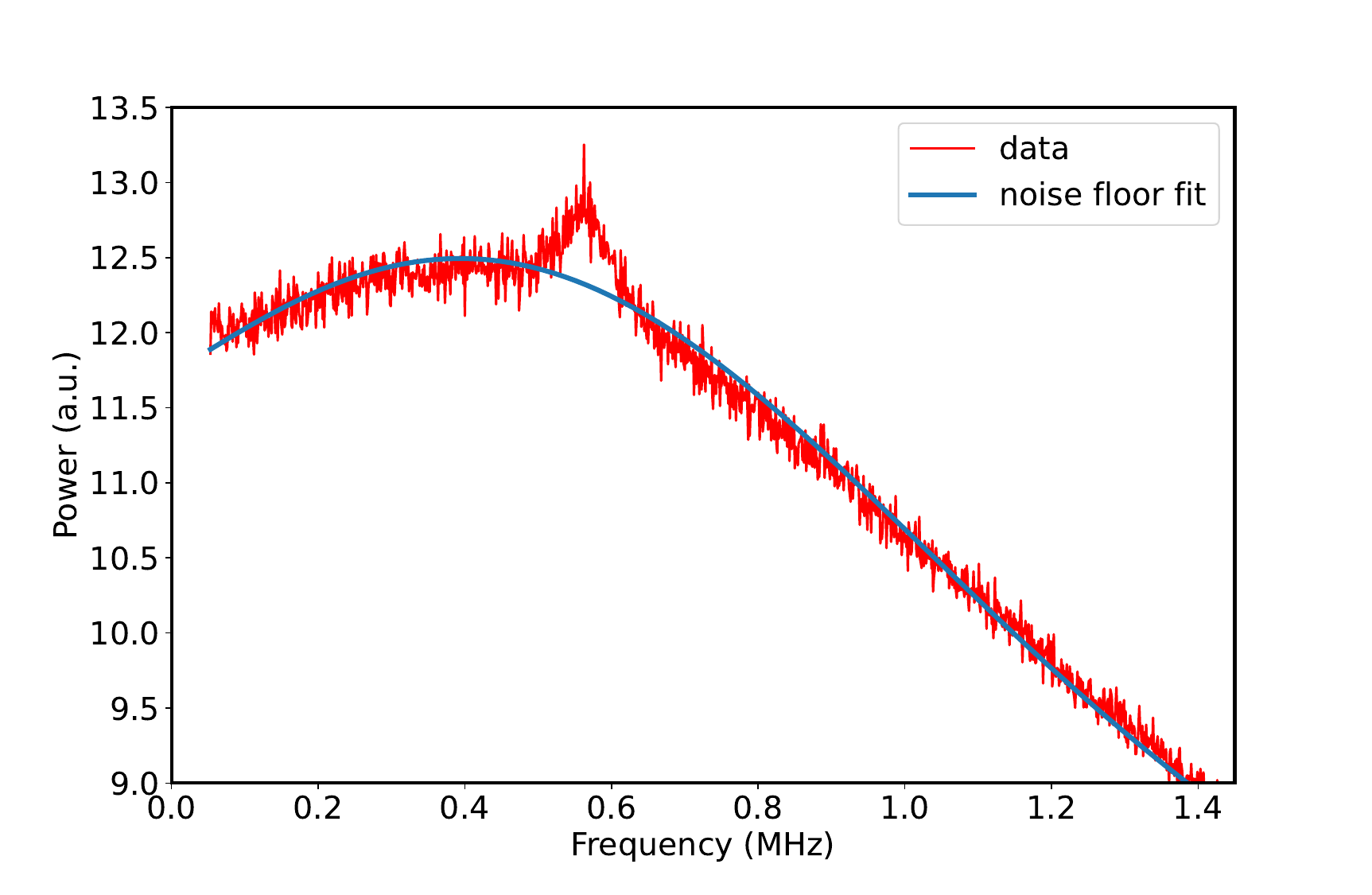}
\caption{Example of a raw PSD trace with 0.22 photons for the weak measurement, which is processed to be a trace in Fig.~\ref{fig:PSDs}. The red line represents the PSD trace obtained from the FFT of the time trace and the blue line represents the fitting to the JPA-dominated noise floor using a polynomial function.
}
\label{fig:raw_FFT}
\end{figure}

With the time traces acquired in the weak measurement regime, the PSD trace is obtained by averaging multiple fast Fourier transforms (FFTs). Each FFT is taken over a 250-ms time trace.  To average enough FFTs for a high-fidelity spectral density, we require less than five minutes' worth of data.  This is a vast improvement over the tens of hours of averaged data required in a previous work~\cite{palacios2010experimental}. Fig. \ref{fig:raw_FFT} is an example of an unprocessed PSD trace. By fitting the noise floor to a polynomial function, we subtract the JPA-response-dominated noise floor and get the PSD of Rabi oscillations under weak measurement in Fig.~\ref{fig:PSDs} in the main text.

\subsection{\label{sec:calibration_driveamp_freqdomain}Rabi oscillation frequencies in the weak measurement regime}
\begin{figure}
\centering
\includegraphics[width=1\columnwidth]{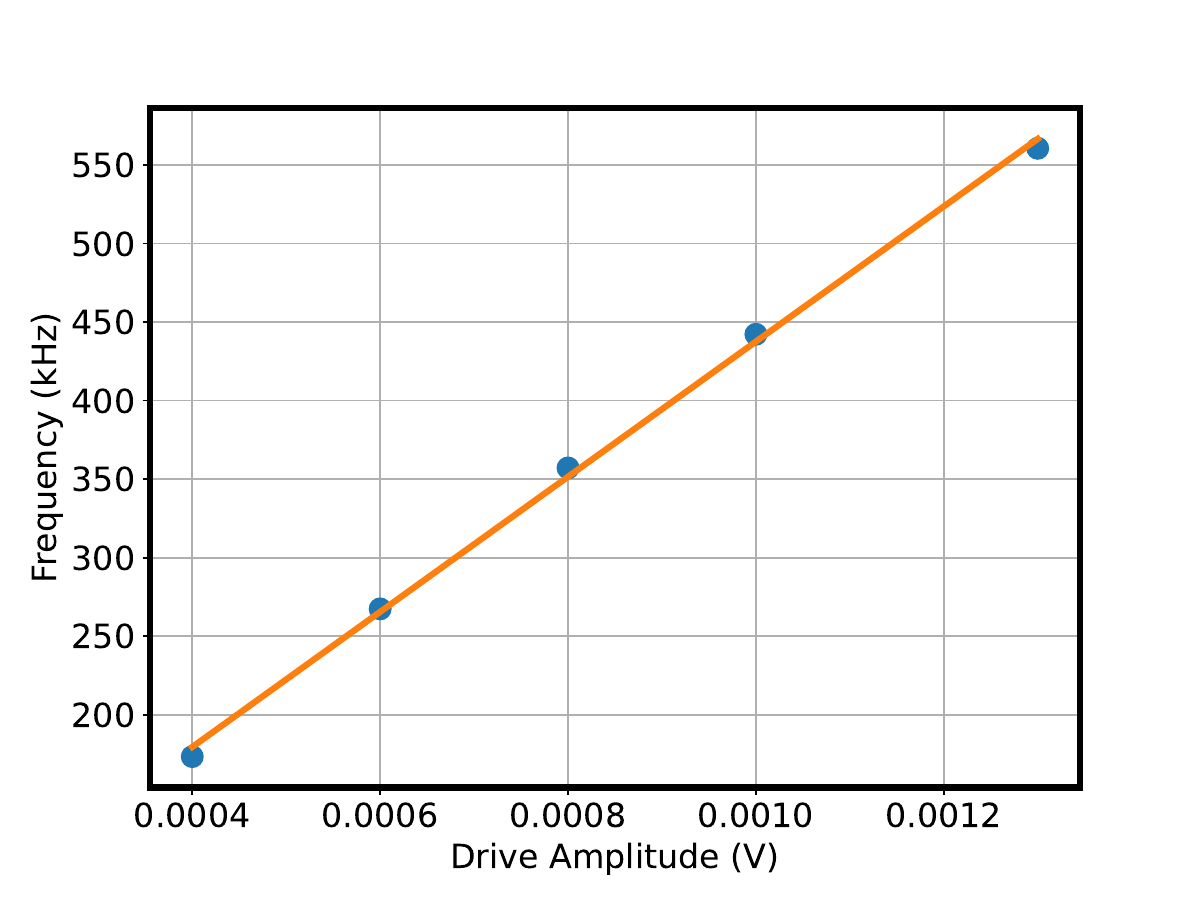}
\caption{Rabi oscillation frequencies extracted from the Lorentzian fit in the weak measurement PSDs in Fig.~\ref{fig:PSDs} for 0.28 readout resonator photon occupation.
}
\label{fig:PSD_freqs}
\end{figure}

By fitting the Lorentzian function to the PSD traces corresponding to a cavity occupation of 0.28 in Fig.~\ref{fig:PSDs} in the main text, the oscillation frequencies of the cavity output signal are obtained. We plot the frequencies versus the qubit drive amplitudes in Fig. \ref{fig:PSD_freqs} and find a linear relation. The linear fit returns a slope the same as the slope in the qubit drive amplitude calibration in Fig. \ref{fig:2D_rabi}.

\subsection{\label{sec:calibration}Photon-number calibration}
\begin{figure}[ht!]
\centering
\hspace*{-0.3in}
\includegraphics[width=1.2\columnwidth]{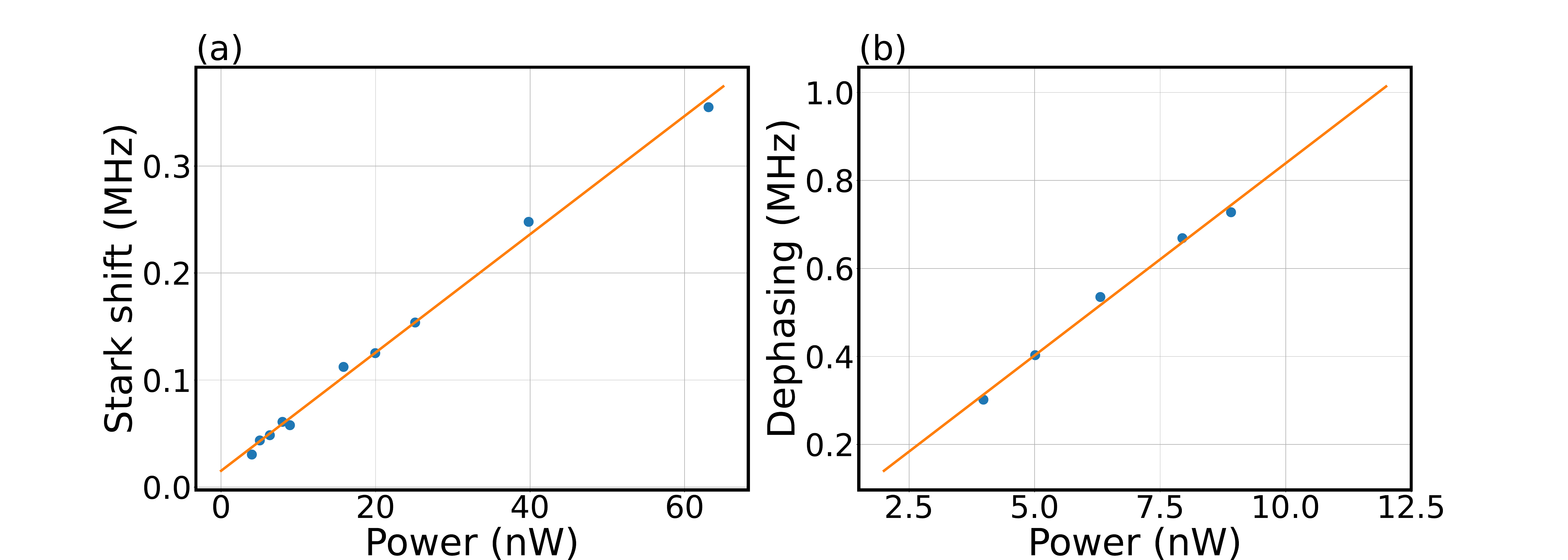}
\caption{(a) ac Stark shift versus cavity drive power; (b) measurement-induced dephasing rate versus cavity drive power.
}
\label{fig:photon_number}
\end{figure}

The dispersive shift $2\chi$ of the readout cavity frequency between the qubit's ground and excited states is determined using pulsed cavity spectroscopy. The pulsed cavity spectroscopy measures the cavity transmission spectrum with and without a $\pi$ pulse. The result in Fig. \ref{fig:photon_number} shows that $\chi/2\pi$ is 340 kHz.

The average intracavity photon number is calibrated with a modified Ramsey measurement, where the free evolution between the two $\pi/2$ pulses is replaced with a square pulse into the readout resonator. This measurement can simultaneously give the ac Stark shift $(\Delta_{\mathrm{ac}} = \chi\bar{n})$ and measurement-induced dephasing rate $(\Gamma=4\chi^2\bar{n}/\kappa)$ of the qubit \cite{Vijay2012}. By sweeping the power of the square pulse, we are able to independently measure the two proportionality constants of the ac Stark shift and measurement-induced dephasing on the power of the square pulsed injected into the cavity. The ratio of the two proportionality constants from the equations is $\kappa/(4\chi)$, and the result from the measurements shown in Fig.~\ref{fig:photon_number} gives a ratio of $0.775 \kappa/(4\chi)$, due to fitting errors. With the verification, we extrapolate the ac Stark shift line to calibrate the average photon number for a given input cavity power. 

\subsection{\label{sec:pre-processing}Preprocessing for the signal of $\langle a + a^{\dagger}\rangle_c$($t$)}
We first rotate the complex data arising from the IQ mixer at the peak frequency in the PSD of each Rabi oscillation to find the phase quadrature that maximizes the signal-to-noise ratio of the oscillation in one quadrature, and the other quadrature is neglected.
The amplitude of the data is then shifted by subtracting the mean value.

\subsection{Data processing for the histograms of oscillatory clock signals}
The inverse Gaussian function used to fit the distribution of the oscillatory clock signal period in Fig.~\ref{fig:wald_fits} in the main text is
\begin{equation}
\label{}
    f(x) = \sqrt{\frac{\lambda}{2\pi x^3}} \mathrm{exp} \left[ \frac{-\lambda (x- \mu)^2}{2\mu^2x} \right] \,\, ,
\end{equation}
where $\mu$ is the mean and $\lambda$ is the spread parameter of the distribution, which are respectively $\pi/\Omega$ and $\pi/\Gamma$ according to Eq.~(\ref{eq:Wald}).

\subsection{Setup}\label{appen:expsetup}
The control and readout microwave pulses from room-temperature instruments are attenuated inside the refrigerator and sent to the qubit chip. The output microwave signal from the chip is isolated by three circulators and amplified by a quantum-noise-limited JPA at the mixing chamber stage and a high-electron-mobility transistor amplifier at the 4-K stage. The output signal is further amplified and down-converted to 25\,MHz with room-temperature instruments. Then a digitizer card inside a computer acquires the signal and digitally down-converts it to a dc signal.

\bibliography{Info-quantum-clock}

\end{document}